\documentclass[twocolumn,preprintnumbers,superscriptaddress]{revtex4-1}
\usepackage{CJK}
\usepackage{mathrsfs}
\usepackage{amssymb}
\usepackage{amsmath}
\usepackage{graphicx}
\usepackage{dcolumn}
\usepackage{bm}
\usepackage{graphicx}
\usepackage{float}
\usepackage[normalem]{ulem}
\usepackage{color}
\usepackage{multirow}
\usepackage{enumerate}

\newcolumntype{d}[1]{Dc{.}{.}{#1}}
\begin{document}
\begin{CJK*}{UTF8}{}

\title{Variational approach for pair optimization in the nucleon pair approximation}
\author{Y. Lei ({\CJKfamily{gbsn}雷杨})}
\email[corresponding author: ]{leiyang19850228@gmail.com}
\affiliation{School of National Defense Science and Technology, Southwest University of Science and Technology, Mianyang 621010, China}
\author{H. Jiang ({\CJKfamily{gbsn}姜慧})}
\affiliation{School of Arts and Sciences, Shanghai Maritime University, Shanghai 201306, China}
\author{S. Pittel}
\affiliation{Bartol Research Institute and Department of Physics and Astronomy, University of Delaware, Newark, Delaware 19716, USA}
\date{\today}

\begin{abstract}
We propose a pair-condensate variational approach (PCV) to determine a set of the most important collective pairs in the description of low-lying states in atomic nuclei. Having available the precise details on these key collective pairs -- their spin, parity, and structure -- can be particularly useful in calculations based on the nucleon-pair approximation (NPA), helping to reduce their uncertainties. In trial calculations for the transitional Ba isotopes, our variational approach describes the evolution of quadrupole-deformation properties similar to Hartree-Fock treatments, while at the same time highlighting the $\gamma$ softness of $^{132}$Ba. Our approach can conclusively determine which collective pairs are critical for obtaining the lowest possible yrast, quasi-beta, quasi-gamma bands, producing both the level structure of these bands and related B(E2) values in reasonable consistency with experiment. These trial calculations suggest that with our PCV approach the NPA can be meaningfully applied to transitional nuclei with a wide spectrum of shapes. We also show that while neutron negative-parity pairs could in principle have an important impact on backbending in $^{132}$Ba, they are not favored for this nucleus.
\end{abstract}
\pacs{XXX}
\maketitle
\end{CJK*}

\section{Introduction}\label{sec-int} Shell-model studies of atomic nuclei \cite{sm-1, sm-2} must typically confront two key issues: knowledge of the nucleon-nucleon interactions and the complexity of the quantum many-body problem. To handle the latter problem, the nucleon pair approximation (NPA) \cite{npa-phys-rep} has been proposed as an efficient shell-model truncation scheme. Inspired by studies of the shell-model foundation of the interacting boson model (IBM) \cite{rev-ibm}, the NPA usually adopts positive-parity $S$ and $D$ pairs with angular momentum $L=0$ and 2, respectively, in analogy with the $s$ and $d$ bosons of the IBM \cite{npa-cal-sd1,npa-cal-sd2,npa-cal-sd4,npa-cal-sd6}. With a few such collective pairs (and the accompanying reduced model space), the resulting wave-functions provide a clear picture of the key components of low-lying nuclear states in a shell-model framework. The NPA can be especially useful for the description of heavy or medium-heavy O(6) nuclei in the transition region between spherical and well-deformed shapes. These nuclei are normally $\gamma$-soft with a large shape uncertainty, and thus are difficult to be interpreted with conventional mean-field theories and very difficult to describe with non-truncated shell-model calculations because of the very large model spaces required.

In addition to the usual $SD$ pairs, other pairs, e.g., a $G$ pair with angular momentum $L=4$ and positive parity \cite{npa-cal-other1} or pairs with negative parity \cite{ npa-cal-other3}, can significantly improve NPA results, especially for states with higher angular momentum. Of course, more pairs means larger model spaces, eventually making an NPA description comparable to a full shell-model treatment \cite{npa-validity-1, npa-validity-2,npa-validity-3}. However, arbitrarily increasing the number of collective pairs violates the original intent of the NPA, viz., to simplify the shell-model description of low-lying states with a highly limited number of collective pairs. It is crucial, therefore, to identify a priori and self-consistently the key collective pairs in advance of an NPA calculation.

Moreover, without an a priori identification of the key collective pairs, there can be some ambiguity about the NPA wavefunctions. For example, consider the $I=10$ backbend in the $^{132}$Ba yrast band \cite{lifetime,gfactor1,gfactor2}. Both $(\nu h_{11/2})^{-2}$ pairs and negative-parity pairs, together with the usual $SD$ pairs, can produce this backbend, but with different NPA wavefunctions \cite{yoshinaga-ba,lei-ba,cheng-ba}. There is still no consensus as to which set of collective pairs is optimal for the description of this $I=10$ backbend.

In the early 1980s, the Hartree-Fock-Bogolyubov (HFB) approach was often adopted to demonstrate the importance of $SD$ pairs in low-lying states, and correspondingly $sd$ bosons in an IBM framework (see, e.g., Refs. \cite{hfb-1, hfb-2}). Those works inspired us to develop a mean-field (like) approach as a pair-importance guide for the NPA. {\it To best serve the needs of an NPA calculation, such an approach should have four additional features}: \begin{itemize} \item {\it It should not violate particle-number conservation.} In the transitional region, where the NPA is mostly used, the nuclear shape can evolve dramatically from spherical to well deformed. Thus, the NPA wave function should also evolve rapidly with increasing valence particle number. Violation of particle-number conservation mixes the wave function of the nucleus under investigation with those of nearby nuclei, leading therefore to a relatively inaccurate description. \item {\it It should enable the NPA to include, or at least have access to, the complete $\gamma$ degree of freedom.} This is crucial if we wish to use this method to understand $\gamma$ softness, $\gamma$ instability, shape coexistence, and shape mixing in transitional nuclei. \item {\it It should be amenable to a hole representation.} To reduce the complexity of the Hamiltonian and transition matrix-element calculations, the NPA, like the shell model, usually adopts a hole representation of the valence space, when the nucleon number is slightly smaller than the corresponding magic number, e.g., southwest of $(N=126, Z=82)$ and northwest of $(N=82, Z=50)$. In such a case, not only the NPA states but also the trial wave function from which collective pairs are generated should be amenable to a description in terms of hole states. \item {\it Such an approach should be able to clarify the importance of both positive- and negative-parity pairs.} In the $A\sim100$, 132 and 208 regions, the intruder $g_{9/2}$, $h_{11/2}$ and $i_{13/2}$ neutron hole states have relatively low single-hole energies and the opposite parity to the other single-hole states in the same major shell. Thus, it is possible to construct low-energy negative-parity collective pairs, in which a low-energy negative-parity hole state is coupled to another hole state with positive parity to produce a pair of overall negative parity. For example, we have demonstrated in earlier work \cite{lei-n74} the indispensable role of negative-parity pairs in the low-lying states of the $N=74$ isotones using the NPA method. \end{itemize}

With precisely the four features specified above, we propose a Pair-Condensate Variational approach (denoted by PCV herein) to facilitate the approximate treatment of the nuclear many-body problem using the NPA method. As we will see, we do not impose a specific angular momentum on the collective pair in the variation, but instead, optimize the linear combination of all possible collective pairs used in the NPA formalism. As a preliminary test of our proposed method, we apply it to the even Ba isotopes from $A=132\sim 136$. These are typical transitional nuclei, and $^{132}$Ba is supposed to be $\gamma$ soft \cite{yoshinaga-ba}. They are all located near or below the $N=82$ closed shell, and thus are optimally treated in hole representation for their neutron shell-model configurations. Such a test should highlight the first three features listed above. The $^{132}$Ba $I=10$ backbend enables us to assess the relative importance of $(\nu h_{11/2})^{-2}$ pairs and negative-parity pairs, as highlighted in the last feature. Moreover, a phenomenological Hamiltonian was already optimized for $^{132}$Ba in Ref. \cite{yoshinaga-ba}, which should help to mitigate uncertainties in nuclear interaction. Therefore, the Ba isotopes are very well-suited test cases for our variational approach.

The paper is organized as follows. In Sec. II, we detail our variational approach, including the formalism, the optimization algorithm, and its properties and discuss the method of calculating deformation parameters and the associated optimal collective-pair decomposition. In Sec. III, we describe our model space and hamiltonian and then discuss the trial calculations we have carried out for the even Ba isotopes and analyze the results that emerged. Two different variational strategies are considered. One averages overall angular momenta in the system, whereas the other makes use of cranking to isolate on states of specific angular momenta. The importance of the latter approach for the description of high spin states is emphasized. Finally, we summarize our results, conclusions, and further prospects in Sec. IV.

\section{Pair-condensate variation}\label{sec-for} \subsection{Formalism} Our method starts with a collective-pair condensate, originally introduced to construct microscopic wave functions of O(6) nuclei and then subsequently to simplify microscopic Monte Carlo shell model calculation for the $A\sim 132$ region \cite{pair-o6,mcsm}. Our formalism for overlaps and Hamiltonian matrix elements is an extension of that used in Refs. \cite{pair-o6,mcsm}, and is summarized in the Appendix.

The ``collective pair'' mentioned above is defined as: \begin{equation}\label{eq-coll-pair} \begin{aligned} &\Lambda^{\dagger}=\frac{1}{2}\sum_{ij}\lambda_{ij} C^{\dagger}_iC^{\dagger}_j ~,\\ &\Lambda=(\Lambda^{\dagger})^{\dagger}=\frac{1}{2}\sum_{ij}\lambda_{ij} C_jC_i~, \end{aligned} \end{equation} where $C^{\dagger}_i$ and $C^{\dagger}_j$ are single-particle creation operators, the $i$ and $j$ indexes represent all the quantum numbers required to label single-particle states, and the $\lambda_{ij}$ are the structure coefficients of the $\Lambda$ collective pair. We enforce $\lambda_{ij}=-\lambda_{ji}$ to ensure the uniqueness of the $\lambda_{ij}$ coefficient. Thus, all the $\lambda_{ij}$ coefficients can be mapped onto a skew-symmetric matrix $\lambda$ as \begin{equation} \lambda= \left( \begin{array}{cccc} 0&\lambda_{12}&\lambda_{13}&\cdots\\ -\lambda_{12}&0&\lambda_{23}&\cdots\\ -\lambda_{13}&-\lambda_{23}&0&\cdots\\ \cdots&\cdots&\cdots&\cdots \end{array} \right)\\~. \end{equation} Anti-symmetry of the $\lambda$ matrix is the key to speeding up the matrix-element calculation of the collective-pair condensate with the optimized BLAS (Basic Linear Algebra Subprograms) \cite{blas} extensions in the MKL (intel Math Kernel Library) library. Therefore, we always anti-symmetrize the coefficient matrix $\lambda$ by $\lambda\leftarrow \frac{1}{2}(\lambda-\lambda^T)$, if it is not skew-symmetric.

We also note that the collective pair defined above does not have angular momentum, angular-momentum projection on the principal axis, or even parity as good quantum numbers. We call such a collective pair as an ``uncoupled collective pair'' in this paper, to distinguish it from conventional collective pairs labeled with definite angular momentum and parity in the NPA. The uncoupled collective pair includes all two-body configuration degrees of freedom, which enables us to determine the importance of all possible NPA collective pairs in a single unbiased variation.

Unless otherwise noted, an uppercase Greek letter in this paper always denotes an uncoupled collective pair, and the corresponding lowercase letter denotes the structure coefficient matrix of this uncoupled collective pair. For example, $\Gamma$ is an uncoupled collective pair, $\gamma$ is its structure coefficient matrix, and $\gamma_{ij}$ is the structure coefficient for the $C^{\dagger}_iC^{\dagger}_j$ configuration.

The trial wave function of our variation is a condensate of uncoupled collective pairs, namely $\left.(\Lambda^{\dagger})^N|\right\rangle$, where $2N$ is the valence particle/hole number in the model space. The pair structure coefficients, namely $\lambda_{ij}$ in Eq. (\ref{eq-coll-pair}), are taken as variational parameters in the PCV, and initialized randomly at the beginning of the variation. Given an arbitrary shell-model Hamiltonian $H$ with one-body and two-body interactions, the Hamiltonian expectation value can be calculated with the formalism described in the Appendix. Our variation minimizes the Hamiltonian expectation value, as expressed by the condition \begin{equation}\label{eq-min} \delta\left(\frac{\left\langle \left(\Lambda\right)^N H\left(\Lambda^{\dagger}\right)^N\right\rangle}{\left\langle \left(\Lambda\right)^N |\left(\Lambda^{\dagger}\right)^N\right\rangle}\right)=0~. \end{equation} We adopt the Broyden-Fletcher-Goldfarb-Shanno algorithm \cite{bfgs-1, bfgs-2,bfgs-3,bfgs-4} for the PCV variation, which requires the first derivatives of the Hamiltonian expectation value.

In Eq. \ref{eq-diff-detail}, we prove that the first derivative of the matrix element $\langle\hat O\rangle$ along the direction of an arbitrary $\Gamma$ pair reads \begin{equation}\label{eq-diff} \frac{\partial \left\langle\left(\Lambda\right)^{N}\hat O(\Lambda^{\dagger})^N\right\rangle}{\partial \delta_{\parallel\Gamma}}=N\left\langle\Gamma\left(\Lambda\right)^{N-1}\tilde {\mathcal{O}}(\Lambda^{\dagger})^N\right\rangle~, \end{equation} where $\hat O$ is an arbitrary linear operator, and $\tilde {\mathcal{O}}=\hat O+\hat O^{\dagger}$. With Eq. (\ref{eq-diff}), we can express first derivatives of Hamiltonian expectation value analytically as in Eqs. (\ref{eq-diff-ove}), (\ref{eq-diff-q}) and (\ref{eq-diff-pp}).

\subsection{Properties of variation}\label{subsec-pro}

Here, we note that the formalism presented in the Appendix does not introduce recursion, and thus a code based on it has polynomial time complexity. The NPA formalism \cite{npa-for-1, npa-for-2}, on the other hand, involves recursion with exponential (or even more aggressive) time complexity. Therefore, as we increase the valence particle number, the computational time of our pair-condensate variation increases more slowly than a conventional NPA code, thus making it a burden for an NPA calculation that incorporates it.

\begin{figure} \includegraphics[width=0.4\textwidth]{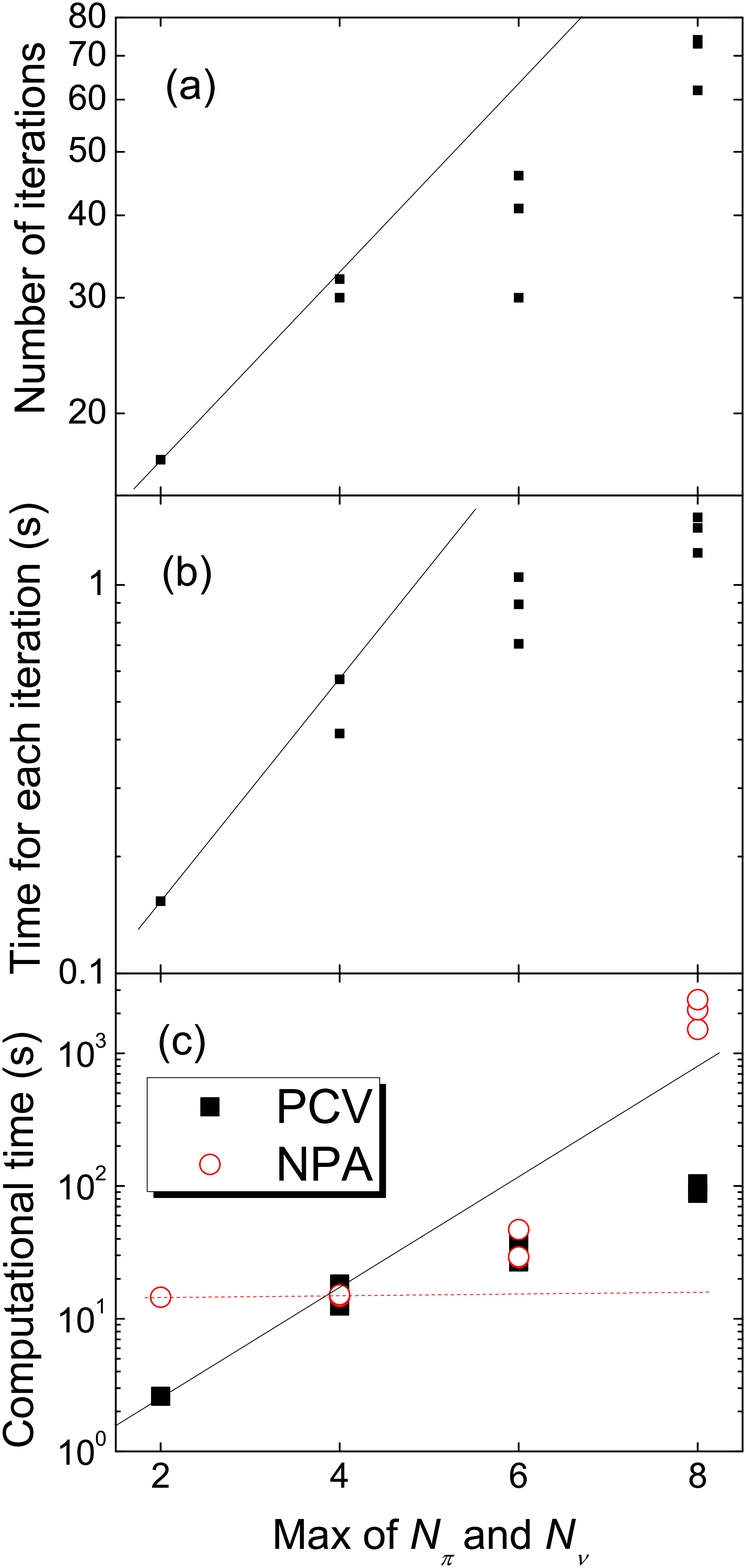} \caption{(Color online) Computational cost of our pair-condensate variation for nuclei in the northwest region of $(N=82,Z=50)$ against the max of valence-proton number and valence-neutron-hole number. The calculational details are described in the text. The straight lines schematically represent an exponential trend. }\label{fig-com} \end{figure}

To illustrate the computational cost of our variational approach, we perform a series of variational calculations for nuclei in the northwest region of $(N=82, Z=50)$ with $\leq 8$ valence protons and $\leq 8$ neutron holes. The single-particle (hole) space and Hamiltonian parameters are specified in Sec. \ref{sec-tri} and we do not go into details on them here. The calculations are performed on a general PC platform with i5-8500 CPU @ 3.00GHz. We find that the computational cost is roughly determined by the maximum of the valence-proton number and the valence-neutron-hole number. Therefore, we plot the iteration number required for convergence and the average computational time of each iteration against the maximum of valence-nucleon numbers in Fig. \ref{fig-com} (a) and (b), respectively. Both computational costs grow more slowly than exponential as increasing valence-nucleon number. Therefore, we expect the total computational time of our variational approach to also grow more slowly than exponential. To illustrate this, we perform NPA calculations with only $SD$ pairs in the same nuclear region and show their computational time in Fig. \ref{fig-com} (c). The computational cost of the NPA has faster growth than exponential, and thus than that of our variational approach, in agreement with the above analysis of the relative time complexity.

Furthermore, with Eq. (\ref{eq-diff}) and the symmetries of the Hamiltonian, it can be proven that variation of the $\left(\Lambda^{\dagger}\right)^N|\rangle$ condensate has three ``self-consistent symmetries'': \begin{enumerate}[1)] \item seniority: if the initial or intermediate $\Lambda$ has only an $S$ pair component, the variation will not allow $\Lambda$ to develop other non-$S$ pair components.\label{item-sph} \item angular-momentum projection: if the initial or intermediate $\Lambda$ has fixed angular-momentum projection, subsequent iterations of the variation will keep this projection until convergence.\label{item-axi} \item parity: if the initial or intermediate $\Lambda$ is labeled with a certain parity, the variation does not change that parity nor mix it with the other parity.\label{item-par} \end{enumerate} One can perform a variation with spherical and axially-symmetric deformation, by imposing symmetries \ref{item-sph}) and \ref{item-axi}), respectively, on the initial $\Lambda$ pair. In our trial calculations, we also observe that the optimized pair condensate always has no parity mixing, even without imposing any symmetry. Such an observation has not yet been proven mathematically as universal.

\subsection{Deformation parameters and pair decomposition}

As noted in Sec. \ref{sec-int}, our pair-condensate variation should provide the NPA access to the full quadrupole-deformed degrees of freedom, and an a priori quantitative measure of collective-pair importance in low-lying states. Therefore, after the variation, we follow the procedure suggested by Ref. \cite{smmc} to determine the quadrupole deformation parameters of the pair-condensate ground state, and to decompose the uncoupled collective pair $\Lambda$ into a series of collective pairs that can be adopted for use in the NPA.

To calculate the deformation parameters $\beta$ and $\gamma$, we first define the total quadrupole operator as $\hat Q=\hat Q_{\pi}-\hat Q_{\nu}$, where $\hat Q_{\pi}$ and $\hat Q_{\nu}$ are the quadrupole operator for proton and neutron, respectively. Here, a negative sign is introduced before the $\hat Q_{\nu}$ operator, since the valence neutrons of the Ba isotopes occupy hole states in our calculations. In Cartesian coordinates, the quadrupole operator $Q_{ij}=3x_ix_j-r^2\delta_{ij}$, where $i$ and $j$ indices refers the three axes of $X,~Y,~Z$, and $\delta$ is a Kronecker symbol. Then, the $\hat Q$ expectation of the pair condensate $\left.(\Lambda^{\dagger})^N|\right\rangle$ can be mapped into a three-dimensional matrix, with three eigenvalues, $\mathcal{Q}_1<\mathcal{Q}_2<\mathcal{Q}_3$. The $\beta$ and $\gamma$ parameters are related to these eigenvalues by \begin{equation} \begin{aligned} \mathcal{Q}_1&=\sqrt{\frac{2 \pi}{5}}\left[\sqrt{3}\left[Q_2+Q_{-2}\right)-\sqrt{2}Q_0\right] \\ \mathcal{Q}_2&=\sqrt{\frac{2 \pi}{5}}\left[-\sqrt{3}\left(Q_2+Q_{-2}\right)-\sqrt{2}Q_0\right] \\ \mathcal{Q}_3&=2 \sqrt{\frac{4 \pi}{5}}Q_0 ~, \end{aligned} \end{equation} with \begin{equation} \begin{aligned} Q_0&=\frac{3}{2 \pi} \sqrt{\frac{4 \pi}{5}}\left\langle r^{2}\right\rangle \beta \cos \gamma \\ Q_2&=\frac{3}{2 \pi} \sqrt{\frac{4 \pi}{5}}\left\langle r^{2}\right\rangle \frac{\beta}{\sqrt{2}} \sin \gamma\\ Q_{2}&=Q_{-2}. \end{aligned} \end{equation}

The NPA calculations make use of collective pairs with definite parity, angular momentum $L$, and projection on the principal axis $M$, as defined by \begin{eqnarray} A^{L\dagger}_M = \sum_{a\leq b} \beta^{LM}_{ab} A^{L\dag}_M (ab),~ A^{L\dag}_M(ab) = \frac{( C^{\dagger}_a \times C^{\dagger}_b )^{(L)}_M}{\sqrt{1+\delta_{ab}}}, \label{pair-definition} \end{eqnarray} where $C^{\dagger}_a$ and $C^{\dagger}_b$ are single-particle creation operators in spherical basis, $a$ and $b$ represent the three-dimensional harmonic-oscillator quantum numbers as normally denoted by $\{nljm\}$, $\beta^{LM}_{ab}$ is the structure coefficient of the collective pair in the NPA. Thus, an arbitrary uncoupled collective $\Lambda$ pair can be rewritten as a sum over $A^{L\dagger}_M$ pairs. If the structure coefficients $\lambda_{ij}$ is also determined in the spherical basis, then the pair structure coefficients $\beta^{LM}_{ab}$ of these $A^{L\dagger}_M$ pairs are determined as \begin{equation} \beta^{LM}_{ab}=\sqrt{1+\delta_{ab}}\sum_{ij}\delta_{j_aj_i}\delta_{j_bj_j}\langle j_im_i,j_jm_j|LM\rangle\lambda_{ij}, \end{equation} where $\delta$ is a Kronecker symbol, and $\langle j_im_i,j_jm_j|LM\rangle$ is an angular-momentum Clebsch-Gordan Coefficient.

Due to the rotational invariance of the nuclear Hamiltonian, the principal axis for the angular-momentum projection is uncertain, and thus the $\beta^{LM}_{ab}$ coefficients are actually varied as a space rotation. This will lead to a linear dependence of the $\{\beta^{LM}_{a_1b_1},\beta^{LM}_{a_1b_2},\beta^{LM}_{a_1b_3},\cdots\}$ vectors with different $M$ values. (To simplify the following description, such vectors are denoted by $\{\beta^{LM}\}$.) One can introduce a unitary transformation to orthogonalize the $\{\beta^{LM}\}$ vectors into $\{\tilde\beta^{LK}\}$ vectors, so that $\sum_{a\leq b}\tilde\beta^{LK}_{ab}\tilde\beta^{LK^{\prime}}_{ab}\equiv 0$ if $K\neq K^{\prime}$. Then, we adopt the $\{\tilde\beta^{LK}\}$ vector as the pair-structure coefficients of the collective pair with angular momentum $L$ in the NPA calculation to follow.

Here, we note that the $K$ index is introduced to distinguish several linearly independent $\{\tilde\beta^{LK}\}$ vectors. It is {\it not} an angular-momentum projection. Furthermore, if we normalize the pair structure coefficients as $\sum_{i<j}(\lambda_{ij})^2=1$, then it can be readily shown that $\sum_{L,K,a\leq b}(\tilde\beta^{LK}_{ab})^2=1$, so that that the length squared of the $\{\tilde\beta^{LK}\}$ vector corresponds to the weight of the collective pair $A^{L\dagger}$ with $\tilde\beta^{LK}_{ab}$ values as structure coefficients. Namely, it can be taken as a quantitative measure of the importance of the collective pair $A^{L\dagger}$.

\section{Trial calculations for the even barium isotopes}\label{sec-tri}

\subsection{Model space and Hamiltonian}

As in earlier NPA calculations for $^{132}$Ba \cite{yoshinaga-ba,lei-ba,cheng-ba}, we limit the single-particle model space to the orbitals of the 50-82 shell. The single-particle motion of the valence neutrons is described in hole representation so that the wave functions involve fewer creation operators. The corresponding collective pair condensate reads $\left.\left.\left(\Lambda^{\dagger}_{\pi}\right)^3\left(\Lambda^{\dagger}_{\nu}\right)^N\right|\right\rangle$, where $\Lambda_{\pi}$ is a proton uncoupled collective pair, $\Lambda_{\nu}$ is the analogous collective neutron hole pair, and $N$ is the number of neutron hole pairs for the Ba isotope under investigation.

In this work, we adopt the same phenomenological Hamiltonian that was proposed in Ref. \cite{yoshinaga-ba}, namely \begin{widetext} \begin{equation}\label{ham} H=\sum\limits_{\sigma=\pi,~\nu} \left(\sum\limits_j\varepsilon_{j\sigma}\hat{n}_{j\sigma} - \sum\limits_{s=0,2}G_{s\sigma}{\cal P}^{(s)\dag}_{\sigma}\cdot \tilde{\cal P}^{(s)}_{\sigma} - \kappa_{\sigma} \hat Q_{\sigma}\cdot \hat Q_{\sigma} \right) +\kappa_{\pi\nu}\hat Q_{\pi}\cdot \hat Q_{\nu}, \end{equation} \end{widetext} with \begin{equation}\label{operator} \begin{aligned} &{\cal P}^{(0)\dagger}= \sum\limits_{a} \frac{\sqrt{2j_{a}+1}}{2}(C_{a}^{\dagger} \times C_{a}^{\dagger})^{(0)},\\ &{\cal P}^{(2)\dagger} = \sum\limits_{ab} q(a b) ( C^{\dagger}_{a} \times C^{\dagger}_{b} )^{(2)},\\ &\hat Q = \sum\limits_{ab} q(a b) ( C^{\dagger}_{a} \times \tilde{C}_{b} )^{(2)}~. \end{aligned} \end{equation} In Eq. (\ref{operator}), all operators are written in a spherical basis. Thus, the structure coefficients of the quadrupole operator $\hat Q$ can be expressed as $q(ab) =-\sqrt{\frac{2j_a+1}{5}}\frac{\langle a||r^2Y^2||b\rangle}{r^2_0}$, where $Y^2$ is the rank-2 spherical harmonic and $r_0=\sqrt{\hbar/(m\lambda)}$ is the oscillator parameter. Also $\varepsilon_{j\sigma}$, $G^{(0)}_{\sigma}$, $G^{(2)}_{\sigma}$ are the single-particle energies and strength parameters of the monopole-pairing and quadrupole-pairing interactions between like-nucleons, while $\kappa_{\sigma}$ and $\kappa_{\pi\nu}$ are the strengths of the quadrupole-quadrupole interactions between like particles and between protons and neutrons, respectively. Table \ref{tab-par} lists explicitly the Hamiltonian parameters as proposed in Ref. \cite{yoshinaga-ba}.

\begin{table} \caption{Adopted Hamiltonian parameters in units of MeV. These parameters were originally proposed for the NPA calculation of $^{132}$Ba \cite{yoshinaga-ba}. The upper part of the table presents the single-particle (s.p.) energies; the lower part presents two-body interaction parameters. There are two different sets of two-body interaction parameters, denoted by PAR-1 and PAR-2 for simplicity.}\label{tab-par} \begin{tabular}{cccccccccccccccccccccccccccccccc} \hline\hline s.p. & $s_{1/2}$ & $d_{3/2}$ & $d_{5/2}$ & $g_{7/2}$ & $h_{11/2}$ & & \\ \hline $\varepsilon_{\pi}$ & 2.990 & 2.708 & 0.962 & 0.000 & 2.793 & & \\ $\varepsilon_{\nu}$ & 0.332 & 0.000 & 1.655 & 2.434 & 0.242 & & \\ \hline\hline two-body & $G_{0\pi}$ & $G_{2\pi}$ & $G_{0\nu}$ & $G_{2\nu}$ & $\kappa_{\pi}$ & $\kappa_{\nu}$ & $\kappa_{\pi\nu}$ \\ \hline PAR-1 & 0.130 & 0.030 & 0.130 & 0.026 & 0.045 & 0.065 & 0.070 \\ PAR-2 & 0.170 & 0.040 & 0.150 & 0.026 & 0.030 & 0.100 & 0.080 \\ \hline\hline \end{tabular} \end{table}

The PAR-1 and PAR-2 Hamiltonian parameters listed in Table \ref{tab-par} have both been used in calculations that reproduce the yrast level scheme of $^{132}$Ba, including its $I=10$ backbend, within the NPA framework. However, they included different sets of collective pairs to achieve such reproduction: 1) PAR-1 included $L^{\pi}=5^-$ and $L^{\pi}=6^-$ collective pairs \cite{lei-ba} in addition to $SD$ pairs, 2) PAR-2 included so-called $H$ pairs based on the $(\nu h_{11/2})^{-2}$ configuration and even angular momentum from $L=0$ to $10$ \cite{yoshinaga-ba} in addition to $SD$ pairs. It has been shown possible to find a unified Hamiltonian that reproduces the backbend in $^{132}$Ba with each set of collective pairs \cite{cheng-ba}. We note here that the PAR-1 and -2 parameters are optimized only for $^{132}$Ba. There may be some difficulty therefore in their ability to reproduce the experimental data for $^{134}$Ba and $^{136}$Ba.

\subsection{Results of the pair-condensate variation}

\begin{table*} \caption{Minimum energies, deformation parameters and collective-pair weights of the optimized collective-pair condensate (PCV) for the various even Ba isotopes considered. The adopted Hamiltonian parameters are listed in Table \ref{tab-par}. The minimum energy is presented in units of MeV. Collective pairs with weights less than 0.1 are omitted here. Here, we list two sets of PCV results with $\omega_X=0$ and $\omega_{\rm B}$. $\omega_X=0$ corresponds to the global minima without cranking, which shall provide collective pairs supposedly important for the major structure of the yrast band. $\omega_X=\omega_{\rm B}$ corresponds to the sudden change of moment of inertia, namely the backbend, as highlighted with blue dotted line in Fig. \ref{fig-omega-e}, where the $\omega_{\rm B}$ values are specified, and the subscript ``$_{\rm B}$'' is the abbreviation of ``backbend''. Thus, the collective pairs with $\omega_X=\omega_{\rm B}$ are supposed to be responsible to induce a backbend.}\label{tab-min} \begin{tabular}{c|c|cccccccccccccccccccccccccccccc} \hline\hline \multicolumn{2}{c}{} & \multicolumn{5}{c}{$\omega_X=0$} & & \multicolumn{5}{c}{$\omega_X=\omega_{\rm B}$} \\ \hline \multirow{11}*{\huge $^{132}$Ba} & \multirow{6}*{PAR-1} & $E_{\rm min}$ & -12.334 & & & & & $E_{\rm min}$ & -9.130 & & & \\ & & $\beta$ & 0.106 & & $\gamma$ & $<1^{\circ}$ & & $\beta$ & 0.093 & & $\gamma$ & $<1^{\circ}$ \\ \cline{3-4}\cline{6-7}\cline{9-10}\cline{12-13} & & \multicolumn{2}{c}{neutron pair weight} & & \multicolumn{2}{c}{proton pair weight} & & \multicolumn{2}{c}{neutron pair weight} & & \multicolumn{2}{c}{proton pair weight} \\ \cline{3-4}\cline{6-7}\cline{9-10}\cline{12-13} & & $L^{\pi}=0^+$ ($S$) & 0.463 & & $L^{\pi}=0^+$ ($S$) & 0.447 & & $L^{\pi}=10^+$ ($H$) & $>0.999$ & & $L^{\pi}=2^+$ ($D$) & 0.273 \\ & & $L^{\pi}=2^+$ ($D$) & 0.453 & & $L^{\pi}=2^+$ ($D$) & 0.489 & & & & & $L^{\pi}=4^+$ ($G$) & 0.399 \\ & & & & & & & & & & & $L^{\pi}=6^+$ ($\mathcal{I}$) & 0.320 \\ \cline{2-13} & \multirow{5}*{PAR-2} & $E_{\rm min}$ & -15.418 & & & & & $E_{\rm min}$ & -11.055 & & & \\ & & $\beta$ & 0.102 & & $\gamma$ & 16$^{\circ}$ & & $\beta$ & 0.083 & & $\gamma$ & 31$^{\circ}$ \\ \cline{3-4}\cline{6-7}\cline{9-10}\cline{12-13} & & \multicolumn{2}{c}{neutron pair weight} & & \multicolumn{2}{c}{proton pair weight} & & \multicolumn{2}{c}{neutron pair weight} & & \multicolumn{2}{c}{proton pair weight} \\ \cline{3-4}\cline{6-7}\cline{9-10}\cline{12-13} & & $L^{\pi}=0^+$ ($S$) & 0.428 & & $L^{\pi}=0^+$ ($S$) & 0.520 & & $L^{\pi}=10^+$ ($H$) & 0.945 & & $L^{\pi}=4^+$ ($G$) & 0.112 \\ & & $L^{\pi}=2^+$ ($D$) & 0.449 & & $L^{\pi}=2^+$ ($D$) & 0.444 & & & & & $L^{\pi}=6^+$ ($\mathcal{I}$) & 0.819 \\ \hline \multirow{10}*{\huge $^{134}$Ba} & \multirow{5}*{PAR-1} & $E_{\rm min}$ & -9.355 & & & & & $E_{\rm min}$ & -5.582 & & & \\ & & $\beta$ & 0.079 & & $\gamma$ & $<1^{\circ}$ & & $\beta$ & 0.049 & & $\gamma$ & $<1^{\circ}$ \\ \cline{3-4}\cline{6-7}\cline{9-10}\cline{12-13} & & \multicolumn{2}{c}{neutron pair weight} & & \multicolumn{2}{c}{proton pair weight} & & \multicolumn{2}{c}{neutron pair weight} & & \multicolumn{2}{c}{proton pair weight} \\ \cline{3-4}\cline{6-7}\cline{9-10}\cline{12-13} & & $L^{\pi}=0^+$ ($S$) & 0.497 & & $L^{\pi}=0^+$ ($S$) & 0.530 & & $L^{\pi}=10^+$ ($H$) & $>0.999$ & & $L^{\pi}=6^+$ ($\mathcal{I}$) & 0.988 \\ & & $L^{\pi}=2^+$ ($D$) & 0.440 & & $L^{\pi}=2^+$ ($D$) & 0.433 & & & & & & \\ \cline{2-13} & \multirow{5}*{PAR-2} & $E_{\rm min}$ & -11.744 & & & & & $E_{\rm min}$ & -7.783 & & & \\ & & $\beta$ & 0.072 & & $\gamma$ & 19$^{\circ}$ & & $\beta$ & 0.056 & & $\gamma$ & 26$^{\circ}$ \\ \cline{3-4}\cline{6-7}\cline{9-10}\cline{12-13} & & \multicolumn{2}{c}{neutron pair weight} & & \multicolumn{2}{c}{proton pair weight} & & \multicolumn{2}{c}{neutron pair weight} & & \multicolumn{2}{c}{proton pair weight} \\ \cline{3-4}\cline{6-7}\cline{9-10}\cline{12-13} & & $L^{\pi}=0^+$ ($S$) & 0.474 & & $L^{\pi}=0^+$ ($S$) & 0.657 & & $L^{\pi}=10^+$ ($H$) & 0.967 & & $L^{\pi}=6^+$ ($\mathcal{I}$) & 0.968 \\ & & $L^{\pi}=2^+$ ($D$) & 0.429 & & $L^{\pi}=2^+$ ($D$) & 0.330 & & & & & & \\ \hline \multirow{10}*{\huge $^{136}$Ba} & \multirow{5}*{PAR-1} & $E_{\rm min}$ & -6.242 & & & & & $E_{\rm min}$ & -2.502 & & & \\ & & $\beta$ & 0.040 & & $\gamma$ & 20$^{\circ}$ & & $\beta$ & 0.032 & & $\gamma$ & $<1^{\circ}$ \\ \cline{3-4}\cline{6-7}\cline{9-10}\cline{12-13} & & \multicolumn{2}{c}{neutron pair weight} & & \multicolumn{2}{c}{proton pair weight} & & \multicolumn{2}{c}{neutron pair weight} & & \multicolumn{2}{c}{proton pair weight} \\ \cline{3-4}\cline{6-7}\cline{9-10}\cline{12-13} & & $L^{\pi}=0^+$ ($S$) & 0.642 & & $L^{\pi}=0^+$ ($S$) & 0.786 & & $L^{\pi}=10^+$ ($H$) & $>0.999$ & & $L^{\pi}=6^+$ ($\mathcal{I}$) & $>0.999$ \\ & & $L^{\pi}=2^+$ ($D$) & 0.349 & & $L^{\pi}=2^+$ ($D$) & 0.210 & & & & & & \\ \cline{2-13} & \multirow{5}*{PAR-2} & $E_{\rm min}$ & -7.806 & & & & & $E_{\rm min}$ & -3.063 & & & \\ & & $\beta$ & 0.031 & & $\gamma$ & $>59^{\circ}$ & & $\beta$ & 0.032 & & $\gamma$ & $<1^{\circ}$ \\ \cline{3-4}\cline{6-7}\cline{9-10}\cline{12-13} & & \multicolumn{2}{c}{neutron pair weight} & & \multicolumn{2}{c}{proton pair weight} & & \multicolumn{2}{c}{neutron pair weight} & & \multicolumn{2}{c}{proton pair weight} \\ \cline{3-4}\cline{6-7}\cline{9-10}\cline{12-13} & & $L^{\pi}=0^+$ ($S$) & 0.701 & & $L^{\pi}=0^+$ ($S$) & 0.911 & & $L^{\pi}=10^+$ ($H$) & $>0.999$ & & $L^{\pi}=6^+$ ($\mathcal{I}$) & 0.946 \\ & & $L^{\pi}=2^+$ ($D$) & 0.296 & & & & & & & & \\ \hline\hline \end{tabular} \end{table*}

Using the parameters listed in Table \ref{tab-par}, we performed PCV calculations for the various Ba isotopes considered in this work. The variational results are listed in Table \ref{tab-min}. For $^{132}$Ba, the PAR-1 and PAR-2 parameters both favor $\beta$ values $\sim 0.1$. With fewer valence neutron holes, the favored $\beta$ value decreases to $\sim 0.03$ as we approach the $N=82$ closed shell, as expected. For the $\gamma$ parameter, the PAR-1 Hamiltonian drives the nuclear shape from prolate to triaxial deformation ($0^{\circ}\rightarrow 20^{\circ}$), whereas the PAR-2 Hamiltonian drives it from triaxial to oblate ($\sim 20^{\circ}\rightarrow 60^{\circ}$).

We also performed Hartree-Fock (HF) calculations for the Ba isotopes using the same Hamiltonians and list the results in Table \ref{tb-hf-min}. The HF method provides a somewhat similar shape evolution as arose in the PVC calculations. As we approach the $N=82$ closed shell, the favored $\beta$ decreases from $\sim 0.1$ to $\sim 0.04$, and the favored $\gamma$ increases towards the oblate ($\rightarrow 60^{\circ}$) value, albeit with sizable differences between the results for PAR-1 and PAR-2. The general similarity in results suggests that the PCV approach can reasonably access the full range of quadrupole degrees of freedom. We also note that PCV always produces lower minimum energy than HF by $\sim 1.5$ MeV, reflecting the gain in energy resulting from the pair correlations introduced in the PCV method.

\begin{table} \caption{Minimum energies and deformation parameters from HF calculations for the various even Ba isotopes considered. The adopted Hamiltonian parameters are listed in Table \ref{tab-par}. The minimum energy is presented in units of MeV.}\label{tb-hf-min} \begin{tabular}{cccccccccccccccccccccccccccccccc} \hline\hline & && $E_{\rm min}$ && $\beta$ && $\gamma$ \\ \hline \multirow{2}*{\large $^{132}$Ba} & PAR-1 && -11.049 && 0.113 && $<1^{\circ}$ \\ & PAR-2 && -14.007 && 0.116 && $13^{\circ}$ \\ \hline \multirow{2}*{ \large $^{134}$Ba} & PAR-1 && -7.878 && 0.093 && $15^{\circ}$ \\ & PAR-2 && -9.641 && 0.093 && $16^{\circ}$ \\ \hline \multirow{2}*{\large $^{136}$Ba} & PAR-1 && -4.604 && 0.045 && $36^{\circ}$ \\ & PAR-2 && -5.643 && 0.044 && $37^{\circ}$ \\ \hline\hline \end{tabular} \end{table}

Using our approach, we can also explore the issue of $\gamma$ softness in $^{132}$Ba. We perform shape-constrained variations across the minimum and along the $\gamma$ direction for this nucleus, using both the HF and PCV approaches. In the HF treatment, the shape constraint is enforced by introducing two Lagrange multipliers, $\lambda_0$ and $\lambda_2$, in the Hamiltonian according to \begin{equation}\label{eq-lin-con} H^{\prime}=H-\lambda_0 r^2Y^2_0-\lambda_2 r^2Y^2_2, \end{equation}
where $Y^2_0$ and $Y^2_2$ are rank-2 spherical harmonics.
This is referred to as a {\it linear constraint} and is widely used in mean-field calculations. However, with such a linear constraint, there is no analytic relation between the Lagrange multipliers ($\lambda_0$ and $\lambda_2$) on the one hand and the deformation parameters ($\beta$ and $\gamma$) on the other. In an HF calculation with a linear constraint, we must repeatedly tune $\lambda_0$ and $\lambda_2$ to achieve specific ($\beta,~\gamma$) values along the $\gamma$ direction across the global minimum.

However, the PCV method is computationally more costly than the HF method, and it is therefore not feasible to tune ($\lambda_0,~\lambda_2$) in this way. Instead, when using the PCV method, we adopt a {\it quadratic constraint}, modifying the expectation value of the Hamiltonian according to \begin{equation} \begin{aligned} \left\langle H^{\prime}\right\rangle=&\left\langle H\right\rangle+\mathcal{C}\left\{\left(\left\langle \hat Q_{zz}\right\rangle-\mu_{zz}\right)^2\right.\\ &\left.+\left(\left\langle \hat Q_{xx}\right\rangle-\mu_{xx}\right)^2+\left\langle \hat Q_{xz}\right\rangle^2\right\}, \end{aligned} \end{equation} where $\mathcal{C}$ is a large positive real number (we take $\mathcal{C}=1000$, herein), $\hat Q_{zz}$, $\hat Q_{xx}$ and $\hat Q_{xz}$ is the quadrupole operator in Cartesian coordinates, and $\mu_{zz}$ and $\mu_{xx}$ are related to the desired $\beta$ and $\gamma$ parameters by
\begin{equation}
\begin{aligned}
\mu_{zz}=&\frac{12}{5}\left\langle r^2\right\rangle \beta\cos\gamma\\
\mu_{xx}=&\frac{6}{5}\left\langle r^2\right\rangle \beta\left[\sqrt{3}\sin\gamma-\cos\gamma\right].
\end{aligned}
\end{equation}

\begin{figure} \includegraphics[angle=0,width=0.48\textwidth]{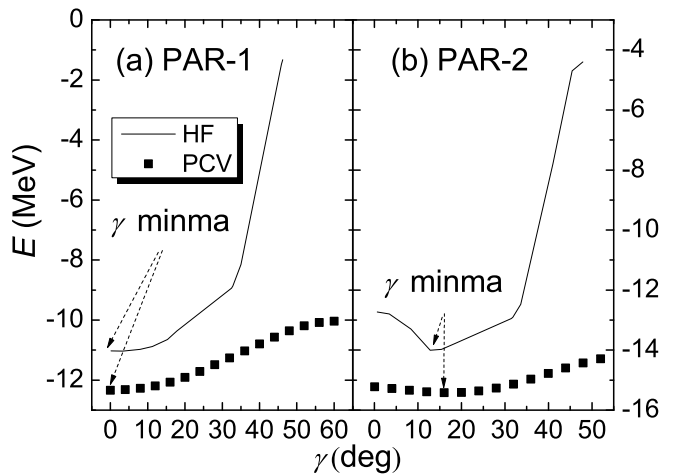} \caption{Potential energy surfaces along the $\gamma$ direction across the minima from shape-constrained HF calculations compared with those from our pair-condensate variational method (PCV). Panels (a) and (b) present the results obtained using the parameters shown in Table \ref{tab-par} for PAR-1 and PAR-2, respectively. The $\gamma$ minima are highlighted here, since some are not very obvious due to the $\gamma$ softness.}\label{fig-pes} \end{figure}

Both the HF and PCV calculations with the associated shape constraints described above provide one-dimensional potential energy surfaces (PES), as illustrated in Fig. \ref{fig-pes}. When comparing the surfaces that derive from the HF and PCV calculations for $^{132}$Ba, we see that the PCV method yields a flatter minimum along the $\gamma$ direction, thereby more effectively establishing $\gamma$ softness for this nucleus. This suggests that our pair condensate method may prove especially useful when trying to obtain an accurate description of $\gamma$-soft nuclei.

From Table \ref{tab-min}, we note that the pair decompositions that derive from the PAR-1 and PAR-2 Hamiltonians are very similar. $SD$ pairs contribute most (over 85\%) of the composition of the optimized $\Lambda$ pairs, reemphasizing the importance of $SD$ pairs in low-lying states that was seen earlier in the previous NPA calculations \cite{npa-phys-rep}.

On the other hand, there is a known $I=10$ backbend in the yrast bands of $^{132-136}$Ba, as reflected by the sudden decrease of $E_I-E_{I-2}$ and B(E2, $I\rightarrow I-2$) at that angular momentum. This backbend at high spins cannot be described solely in terms of $S$ and $D$ pairs, requiring additional higher spin pairs as well \cite{yoshinaga-ba,lei-ba,cheng-ba}.

A frequent approach to deal with a backbend in rotational bands is through the introduction of cranking \cite{in-1,in-2} in the microscopic theory being used \cite{crank-review-1,crank-review-2,crank-review-3}. For example, Ref. \cite{hfb-1} adopted the cranked HFB approximation to study the evolution of $SD$ pairs across a backbend. Similarly, the Cranked Shell Model has been applied extensively to understand the $I=10$ backbend in the Ba isotopes \cite{gfactor2, crank-ba-1,crank-ba-2}. To identify and construct the appropriate collective pairs for the backbend in the PCV approach, we too will introduce cranking, by introducing the cranked Hamiltonian \begin{equation} H_{\rm crank}=H-\omega_XJ_X. \end{equation} Here $\omega_X$ is the angular velocity and $J_X$ is the angular momentum projection on the $X$ axis. Compared with the shape constraint in HF, as illustrated by Eq. (\ref{eq-lin-con}), the minimization of $H_{\rm crank}$ corresponds to a linear constraint of the angular-momentum expectation value ($\langle J_X\rangle$), with $\omega_X$ a Lagrange multiplier chosen to yield a given angular momentum on average. Thus, with an appropriate $\omega_X$, we can identify the key collective pairs for excited states with higher spin, e.g., the $I^{\pi}=10^+$ yrast isomer, which is critical for the $I=10$ backbend.

\begin{figure} \includegraphics[angle=0,width=0.48\textwidth]{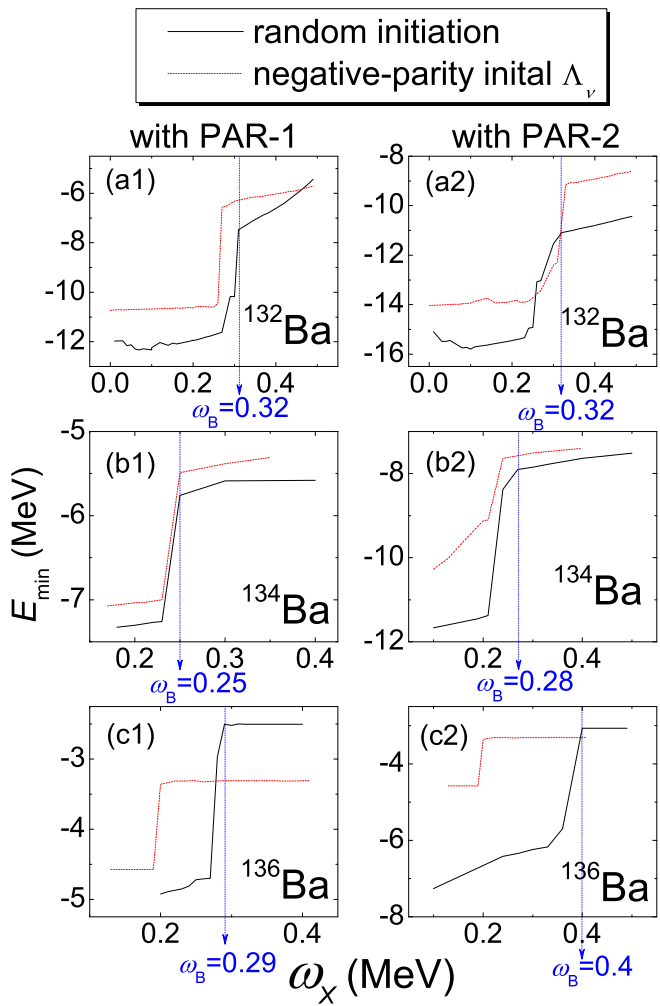} \caption{(Color online) Minimum energy versus angular velocity $\omega_X$ from the cranked PCV variation. The blank line corresponds to a normal variation with random initiation, while the red dashed line corresponds to initiation with negative-parity $\Lambda_{\nu}$. The sudden change of moment of inertia from the normal PCV variation is highlighted by the blue dotted line, where a backbend occurs, and the corresponding angular velocity ($\omega_{\rm B}$) is also specified. The subscript ``$_{\rm B}$'' is an abbreviation for ``backbend''. We extract the collective pairs from $\omega_X=\omega_{\rm B}$ variations as listed in the right column of Table \ref{tab-min}. These should be the pairs that are responsible for the yrast $I=10$ backend of the Ba isotopes. \label{fig-omega-e}} \end{figure}

We plot the evolution of the minimum energy of the pair condensate with increasing angular velocity ($\omega_X$) in Fig. \ref{fig-omega-e} for the various nuclei under investigation. A sharp rise of the minimum energy is seen for all of the nuclei studied around $\omega_X\sim 0.3$, which corresponds to a sudden change of the moment of inertia, i.e., to a backbend. We extract the collective pairs from the optimized pair condensate after the backbend (as highlighted by the blue dotted lines in Fig. \ref{fig-omega-e}). The collective pairs extracted in this way should be important to construct the second rotational band, which crosses the ground band and in doing so induces the backbend.
We also show these collective pairs in Table \ref{tab-min}. For both the PAR-1 and PAR-2 parameter sets, the dominant neutron collective pair is an $H^{L=10}$ pair that arises from the $(\nu h_{11/2})^{-2}$ configuration. The dominant proton collective pair in $^{134}$Ba and $^{136}$Ba is an $L^{\pi}=6^+$ pair, arising from the $(\pi g_{7/2})^{2}$ configuration. In $^{132}Ba$, proton $D$ ($L^{\pi}=2^+$) and $G$ ($L^{\pi}=4^+$) pairs also contribute substantially for the PAR-1 Hamiltonian whereas a proton $G$ pair contributes substantially for the PAR-2 Hamiltonian. We note that the importance of the neutron $H^{L=10}$ pair for the $I=10$ backbend was already emphasized in Ref. \cite{yoshinaga-ba}. The previous CSM calculation also found a band crossing frequency for $(\nu h_{11/2})^{-2}$ alignment in $^{132}$Ba at $\omega_X\sim 0.3$ \cite{gfactor2,crank-ba-1,crank-ba-2}, in agreement with our cranking results.

In Ref. \cite{lei-ba}, it was proposed that the inclusion of neutron negative-parity pairs may be an alternative approach for describing the $I=10$ backbend in $^{132}$Ba. However, the pair-condensate variational results after the backbend suggest otherwise and indeed we will see in the next subsection that even without any negative-parity pairs, the NPA can achieve an excellent description of the $^{132}$Ba backbend. How can we understand why such a pair was able to produce a reasonable description of the backbend in those earlier calculations even though not favored by our cranking treatment? To address this, we impose negative parity on the initial $\Lambda_{\nu}$ pair of our cranking treatment and then perform another variational treatment for the Ba isotopes. According to the parity self-consistent symmetry, namely symmetry \ref{item-par}) described in Subsection \ref{subsec-pro}, such a variation should converge to a negative-parity $\Lambda_{\nu}$ pair. We also plot the minimum energy of that solution against the angular velocity ($\omega_X$) in Fig. \ref{fig-omega-e}. For $^{132}$Ba and $^{134}$Ba, the negative-parity $\Lambda_{\nu}$ pair also produces a sharp rise of the minimum energy much as the positive-parity pair does. For $^{136}$Ba, however, the points at which the sharp rise occurs have a difference in the angular velocity larger than 0.1 MeV.

It requires at least two negative-parity pairs to produce a similar cranking plot to the plot with positive-parity pairs. These negative-parity pairs can provide two $h_{11/2}$ holes, with which the $(\nu h_{11/2})^{-2}$ configuration, i.e., the $H^{L=10}$ pair can be reconstructed. Such reconstruction connects the negative-pair coupling and the $H^{L=10}$ pair and thus explains why negative-parity collective pairs were able to describe the $H^{L=10}$-pair dominant $I=10$ backbend in our previous NPA calculation \cite{lei-ba} for $^{132}$Ba. On the other hand, we also note that cranking with negative-parity $\Lambda_{\nu}$ pairs leads to a higher minimum energy than cranking with positive parity for almost all the $\omega_X$ values. For this reason, negative-parity neutron collective pairs are not favored, even for the PAR-1 parameter set, and thus will not be used in the NPA calculations to follow.

\subsection{NPA calculations using PBCS and PCV pairs}

\begin{figure*} \includegraphics[angle=0,width=1.0\textwidth]{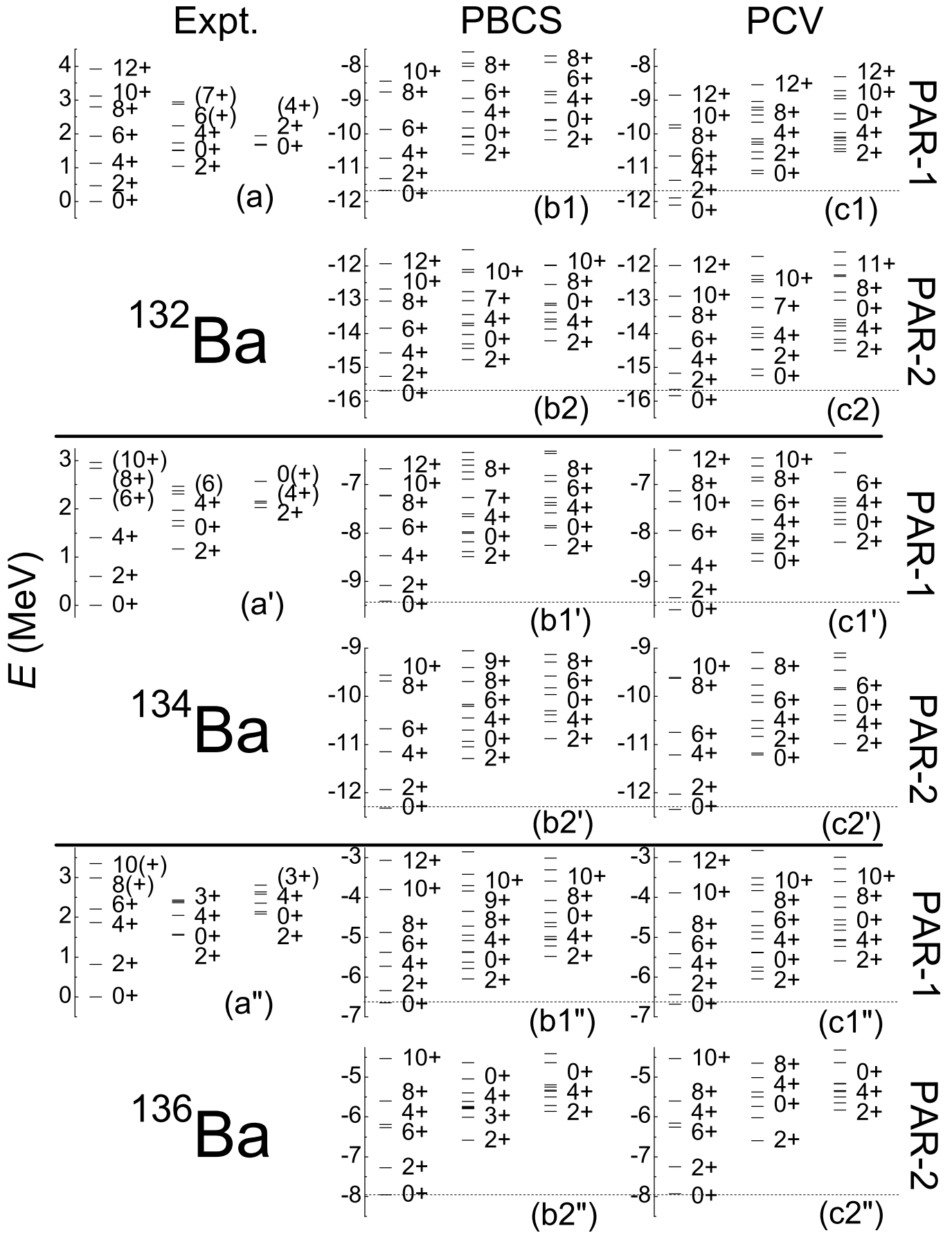} \caption{Level schemes from experiment \cite{ensdf} (labeled ``Expt.") and NPA calculations with the ``PBCS'' and ``PCV'' pair-optimization approaches. Results with the PAR-1 and PAR-2 parameters shown in Table \ref{tab-par} are both presented. The ``PBCS'' calculation for $^{132}$Ba follows Refs. \cite{yoshinaga-ba,lei-ba}. All the other calculations include the collective pairs listed in Table \ref{tab-min}. The horizontal dashed lines highlight the ground states of the ``PBCS'' calculations for a clearer comparison.}\label{fig-spe-all} \end{figure*}

We introduce the collective pairs listed in Table \ref{tab-min} in our NPA calculations with the PAR-1 and PAR-2 Hamiltonians of Eq. (\ref{ham}). The level schemes obtained using the collective pairs from the PCV variational method are plotted in the PCV column of Figs. \ref{fig-spe-all}, including the yrast, quasi-beta, and quasi-gamma bands. We compare the calculated results with the experimental data \cite{ensdf}, as well as with results from conventional NPA calculations in which collective pairs are obtained from a numerical fit to projected BCS wave functions (PBCS) \cite{pbcs}. We note that the PBCS pairs were applied earlier in NPA calculations for $^{132}$Ba \cite{lei-ba,cheng-ba}, so that the $^{132}$Ba results with PBCS presented here are simply a replot of those earlier results. Following Ref. \cite{lei-ba}, our NPA calculation with the PBCS pair for $^{132}$Ba includes the appropriate $SD$ pairs as well as $L^{\pi}=5^-$, and $L^{\pi}=6^-$ pairs when we use the PAR-1 parameters. When we use the PAR-2 parameters, the relevant pairs included are its $SD$ pairs and all the $H^{L=2\sim 10}$ pairs, as in Ref. \cite{yoshinaga-ba}. For $^{134,~136}$Ba, the NPA calculations with PBCS adopt the same collective-pair sets as for PCV (as listed in Table \ref{tab-min}) to permit a meaningful comparison, although they do not have the same pair structures.

We first focus on a comparison between our calculated level schemes and those from experiments. Generally speaking, the NPA calculations with both the PBCS and PCV collective pairs achieve a reasonable agreement with experiments for $^{132}$Ba. However, when we change the valence neutron-hole number, the agreement gradually worsens. In particular, the yrast bands of $^{134}$Ba and $^{136}$Ba resulting from the NPA calculations show somewhat larger moments of inertia than the experimental data, and the calculated quasi-beta and gamma bands are lower in energy than in experiment for these two nuclei. The reason is that our Hamiltonian parameters were optimized for $^{132}$Ba alone \cite{yoshinaga-ba} so that greater disagreement with experiments for the other nuclei is to be expected.

\begin{figure} \includegraphics[angle=0,width=0.48\textwidth]{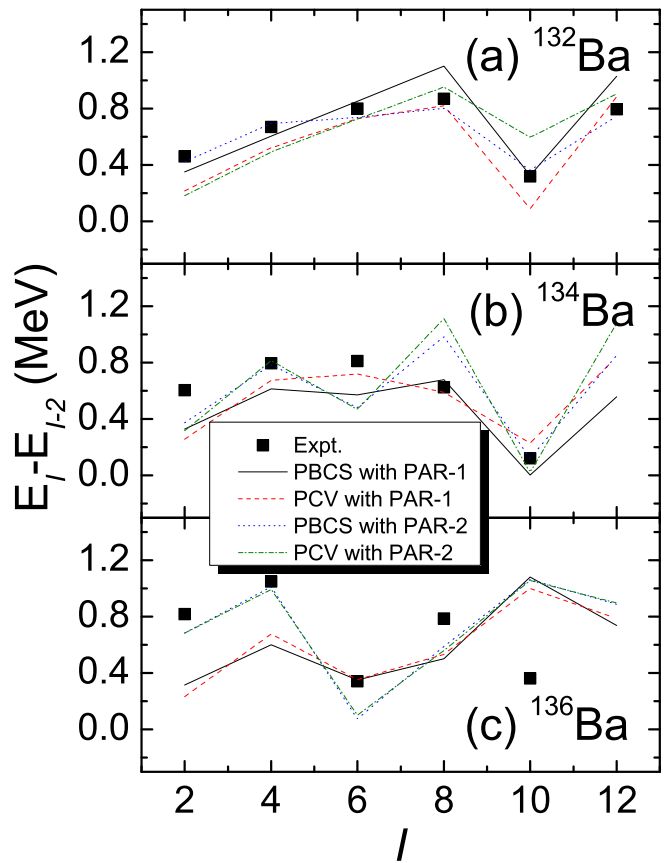} \caption{(Color online) $E_I-E_{I-2}$ in the yrast bands of the even Ba isotopes. The experimental data (Expt.) is from Ref. \cite{ensdf}.}\label{fig-dei} \end{figure}

We would now like to discuss the yrast backbend in a bit more detail. The backbend can be most clearly seen and discussed through an $E_I-E_{I-2}$ .vs. $I$ plot of the yrast band levels, which we present in Fig. \ref{fig-dei}. The experimental data for all three Ba isotopes exhibit an $I=10$ backbend with relatively small $E_{10}-E_8$ values. The NPA calculations reproduce such a backbend for both $^{132}$Ba and $^{134}$Ba but seem to fail for $^{136}$Ba. Instead, we observe in our calculation for this nucleus a sudden decrease of $E_I-E_{I-2}$ for $I=6$ in the yrast band. This is another success of our NPA treatment since experiments also suggest a sudden decrease of $E_6-E_4$, as can be seen in Fig. \ref{fig-dei}(c).

It is important to reiterate here that the variational PCV analysis did not give rise to any negative parity pairs to be included in the NPA calculations, as is evident from Table \ref{tab-min}. Instead, the $I=10$ backbend was produced solely by including the appropriate positive-parity pairs, even for the PAR-1 Hamiltonian. This is different from the earlier explanation of the $I=10$ backbend mechanism in Ref. \cite{lei-ba}, which resulted from an arbitrary choice of collective pairs, as is inherent in the PBCS approach. In contrast, the PCV method provides a well-defined and unambiguous way to choose the dominant pairs, so that we can now finally pin down conclusively the $I=10$ backbend mechanism of $^{132}$Ba.

Perhaps most importantly, both the PAR-1 and PAR-2 calculations have the common feature that the $I=10$ backbend is spontaneously produced (the sudden drop of level space $E_{I}-E_{I-2}$) when we introduce cranking in our pair-condensed variational analysis and then use the resulting pairs in the NPA. (Without cranking, pure $SD$ calculations do not reproduce the $I=10$ backbend \cite{yoshinaga-ba}.) This demonstrates fairly convincingly that the introduction of cranking in our variational analysis is indeed a practical way to improve the NPA when dealing with higher spin states and backbending phenomena.

To further demonstrate the usefulness of the PCV approach, we compare the NPA results that are obtained using the collective pairs from the PBCS and PCV approaches, i.e., the right two columns of Fig. \ref{fig-spe-all}, respectively. For $^{132}$Ba with PAR-1, we see that levels from the PCV calculation (Fig. \ref{fig-spe-all}(c1)) are systematically lower than those from the corresponding PBCS calculation (Fig. \ref{fig-spe-all}(b1)). A reasonable low-energy truncation scheme of the shell model should provide the lowest yrast levels possible since lower energy hints larger overlap between the eigenstates from a truncated subspace and those from the full shell-model space. The comparison just noted suggests that PCV is superior to PBCS for $^{132}$Ba for the PAR-1 parameter set. With PAR-2, as shown in Fig. \ref{fig-spe-all}(b2) and (c2), some improvement of PCV over PBCS can likewise be noted, but the difference is not as dramatic as in the PAR-1 analysis. The less dramatic improvement can be attributed to the fact that the PBCS calculations with PAR-2 have been highly optimized with the $H_{\nu}^{L=10}$ pair in Refs. \cite{yoshinaga-ba}. Since the PBCS with PAR-2 starts from a similar backbend mechanism to the PCV, they should provide similar spectra, and this is illustrated in Fig. \ref{fig-spe-all}(b2) and (c2). However, we note that the $H_{\nu}^{L=10}$ pair had to be artificially introduced in the PBCS pair optimization, while in PCV the pair arises in a self-consistent and a priori way when considering $I=10$ backbend. Here too we see the key benefit provided by the PCV approach.

As we approach the $N=82$ closed shell, the difference between the PBCS and PCV level schemes becomes negligible, although PCV still produces slightly lower levels. For $^{136}$Ba, with $N=80$, the two approaches again provide nearly identical level spectra. As shown in Table \ref{tab-min}, the nuclear deformation becomes very small as we approach the shell closure, and it is precisely for very small deformations where the PBCS method with only an $S$-pair condensate works well.

\begin{figure} \includegraphics[angle=0,width=0.48\textwidth]{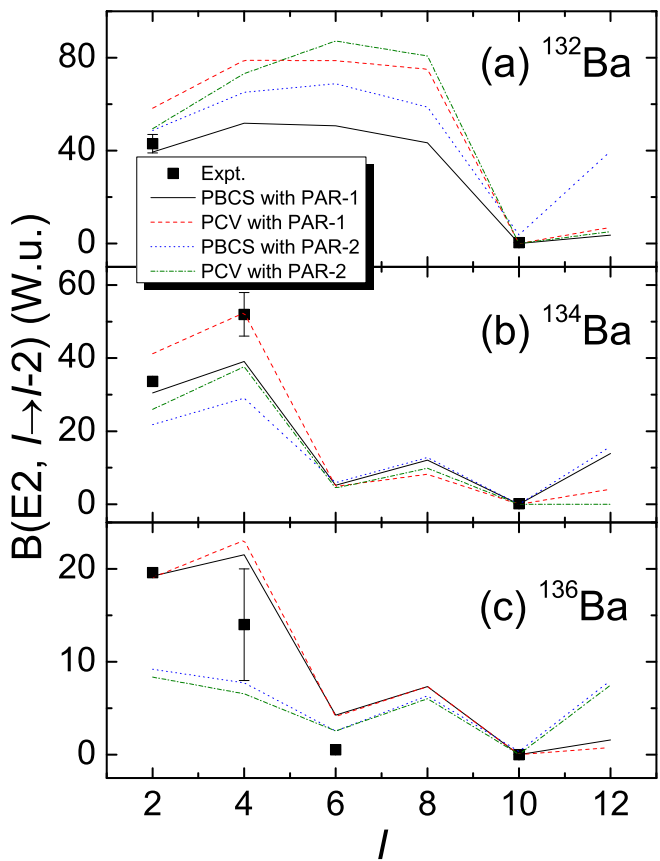} \caption{(Color online) Same as Fig. \ref{fig-dei} except for B(E2, $I\rightarrow I-2$). The effective charges are $e_{\pi}=2e$ and $e_{\nu}=-1e$, following Refs. \cite{yoshinaga-ba,lei-ba}}\label{fig-be2} \end{figure}

As is well known, E2 transition rates can provide a sensitive probe of nuclear structure wave functions. As an example, the $I=10$ backbend in the Ba isotopes is also reflected by the reduced B(E2, $10\rightarrow 8$) values, as can be seen e.g. in Fig. \ref{fig-be2}. This feature is nicely reproduced by our NPA calculations, both with the PBCS and PCV approaches, for all of the Ba isotopes under investigation. In the B(E2) calculations shown in Fig. \ref{fig-be2}, we used effective charges of $e_{\pi}=2e$ and $e_{\nu}=-1e$, following Refs. \cite{yoshinaga-ba,lei-ba}. Thus the NPA success in describing spectral features of the $I=10$ backbend in Fig. \ref{fig-spe-all} ( except for $^{136}$Ba) carries over to E2 properties as well. We should also note the small B(E2, $6\rightarrow 4$) of $^{136}$Ba in Fig. \ref{fig-be2}(c), which is correlated with the small energy gap of $E_6-E_4$ in Fig. \ref{fig-spe-all}(a$^{\prime\prime}$). It seems that a new ``backbend'' around $I=6$ may be occurring with a mechanism similar to that of the $I=10$ backbend. Of course, it is not appropriate to speak about a rotational band in a nearly spherical nucleus like $^{136}$Ba, which is the reason we put the term "backbend" here in quotes. Nevertheless, we can now understand why our calculated spectrum for $^{136}$Ba does not produce an $I=10$ ``backbend'', even though the B(E2) values seem to. The calculated $I=6$ ``backbend'' lowers the $8^+$ state too much to create a small energy gap of $E_{10}-E_8$, but it does not reduce the significant difference between the $8^+$ and $10^+$ wave-functions, as reflected by the small B(E2, $10\rightarrow 8$) value. Thus, we believe that the adopted Hamiltonian parameters provide reasonable wave functions for the {\it full set} of $^{132-136}$Ba nuclei.

In Fig. \ref{fig-be2}, we also note that for $^{132}$Ba the PCV calculation leads to larger B(E2) values than the PBCS calculation before the backbend. This is to be expected since the PBCS method neglects the proton-neutron interaction during the pair optimization, thus suppressing some of the configuration mixing and collectivity that would ensue from these correlations, whereas the PCV calculations include them. For $^{134}$Ba and $^{136}$Ba, the two approaches give rise to similar B(E2) evolution with increasing spin $I$. As for the spectral comparison, the difference between B(E2) values from the PBCS and PCV calculations likewise fades as we approach the $N=82$ shell closure. Thus, the PCV approach can provide a good description of the nuclear collectivity exhibited by deformed nuclei while at the same time reducing to the PBCS approach for spherical nuclei. It can therefore flexibly describe the entire nuclear transition region with a wide spectrum of nuclear shapes, something that was not possible with earlier treatments.

To more fully demonstrate the validity of our proposed pair-condensate variation, it would be useful to evaluate the overlap between NPA wave functions and the corresponding shell-model wave functions, as was done in Refs. \cite{pbcs,val-2,val-3}. Unfortunately, it is not possible to carry out such an analysis for the Ba isotopes at present. Extended verification of our method with smaller model spaces and with a variety of interactions should be explored.

\section{summary}\label{sec-sum} In the present work, we have proposed a pair-condensate variational approach to improve NPA calculations and to define the optimal collective pairs for a description of atomic nuclei. The variational method has particle-number conservation, as needed to accurately describe transitional nuclei, the regime in which the NPA is especially useful. It also includes all the two-body-configuration degrees of freedom. Therefore, it works for asymmetric deformed nuclei and can estimate the possible importance of negative-parity collective pairs. It can also be applied in a hole representation to enable the investigation of transitional regions slightly below magic numbers. We have also proposed three self-consistent symmetries for such a variational analysis.

We have performed trial calculations for the even Ba isotopes in the northwest transitional region above ($N=82,~Z=50$). The proposed variational approach can describe the nuclear shape of the various Ba isotopes we considered and, most importantly, provides an improvement over earlier NPA calculations. In detail, the variational method produces similar nuclear shape evolution to an analogous HF calculation and establishes $\gamma$ softness for the nucleus $^{132}$Ba. This illustrates both the validity of our variational approach and its ability to describe asymmetric deformation. By incorporating cranking in our variational approach, we have been able to demonstrate the key role of the neutron $H^{L=10}$ pair and proton $\mathcal{I}$ pairs in the description of the higher $I$ yrast states. The NPA calculations carried out using the input from our cranked variational approach were able to successfully and self-consistently describe the low-lying level schemes of the even Ba isotopes from $^{132}$Ba through $^{136}$Ba and their $I=10$ backbend, while producing lower energies than earlier NPA calculations \cite{yoshinaga-ba,lei-ba}. We also explained why negative-parity pairs could in principle produce a backbend in $^{132}$Ba but are nevertheless not recommended for an optimal description within the framework of our method.

Finally, we wish to emphasize here that previous NPA calculations (e.g. in Refs.\cite{yoshinaga-ba,lei-ba,cheng-ba}) invariably adopted collective pairs that were blindly chosen, adjusted, and never fully justified. With the pair-condensate variational approach we have proposed in this work, NPA calculations can now be carried out with greater confidence on which collective pairs to include and on their structure. In subsequent work, we will further study the possibility of parity mixture in the optimized pair condensate, extensively verify the validity of our variational approach by wave function overlap analysis, and will then apply the method to more realistic NPA calculations.

\acknowledgements We thank C. W. Johnson for providing an HF code \cite{sherpa} and for helpful suggestions. This work also benefited from intensive discussions with C. F. Jiao and J. M. Yao on the possibility of carrying out an HFB calculation for $^{132}$Ba. We are grateful for the financial support of the Sichuan Science and Technology Program (Grant No. 2019JDRC0017), the Doctoral Program of Southwest University of Science and Technology (Grant No. 18zx7147), and the National Natural Science Foundation of China (Grant No. 11875188).

\appendix* \section{Formalism for pair condensate} Unless specifically noted to the contrary, an uppercase Greek letter in this Appendix always denotes an uncoupled collective pair and the corresponding lower case letter denotes the structure coefficient matrix of this uncoupled collective pair. Thus, for example, if $\Gamma$ is an uncoupled collective pair, then $\gamma$ is its structure coefficient matrix. Furthermore, $\gamma_{ij}$ is the structure coefficient associated with the $C^{\dagger}_iC^{\dagger}_j$ pair configuration or with the $C^{\dagger}_iC_j$ particle-hole configuration.

Now consider the contraction of two uncoupled collective pairs, $\Gamma$ and $\Lambda$, which can be expressed as
\begin{equation} \left[\Gamma,\Lambda^{\dagger}\right]=-\frac{1}{2}tr(q)+Q \end{equation}
where $q$ is a matrix of the form $q=\lambda\gamma$, $tr(q)$ is the trace of the $q$ matrix, and $Q$ is a one-body operator with $q$ as its structure coefficient matrix, viz., $Q=\sum q_{ij}C^{\dagger}_iC_j$ \cite{pair-o6,mcsm}, .

The contraction of a collective pair and an arbitrary one-body operator reads
\begin{equation}
\begin{aligned} \left[\Gamma,Q\right]=\frac{1}{2}\sum_{ijkl}\gamma_{ij}q_{kl}\left[C_jC_i,C^{\dagger}_kC_l\right]=\sum_{ijl}q_{il}\gamma_{ij}C_jC_l.\\ \end{aligned}
\end{equation}
If we let $\Lambda=\left[\Gamma,Q\right]$, then the coefficient matrix of $\Lambda$ is $\lambda=\gamma q+q^T\gamma$, where $q^T$ is the transpose of of matrix $q$. The above two contractions will be frequently used in the formalism to follow.

As in Refs. \cite{pair-o6,mcsm}, the overlap of the pair condensate $(\Lambda^{\dagger})^{N}|\rangle$ is denoted by $I^N$, and reads \begin{equation}\label{eq-ove-0} \begin{aligned} I^N=\langle (\Lambda)^N(\Lambda^{\dagger})^N\rangle=-\frac{1}{2}N\sum_{l=0}^{N-1}tr(\lambda^{2l+2})J^{N-1}_l, \end{aligned} \end{equation} where \begin{equation}\label{eq-J-tensor} J^N_l=\left[\frac{N!}{(N-l)!}\right]^2I^{N-l}. \end{equation}

To further calculate the required Hamiltonian matrix elements, we also need the formalism for three other overlaps, viz., $\left\langle\Gamma_1\left(\Lambda\right)^{N-1}\left(\Lambda^{\dagger}\right)^{N}\right\rangle$, $\left\langle \Gamma_1\Gamma_2\left(\Lambda\right)^{N-2}\left(\Lambda^{\dagger}\right)^{N}\right\rangle$, and $\left\langle \Gamma_1\Gamma_2\Gamma_3\left(\Lambda\right)^{N-3}\left(\Lambda^{\dagger}\right)^{N}\right\rangle$, where $\Gamma_1$, $\Gamma_2$ and $\Gamma_3$ are arbitrary pairs with coefficients matrices $\gamma_1$, $\gamma_2$ and $\gamma_3$, respectively. To simplify our formalism, these three overlaps are denoted by $\left\langle\gamma_1,N\right\rangle$, $\left\langle\gamma_1,\gamma_2,N\right\rangle$, and $\left\langle\gamma_1,\gamma_2,\gamma_3,N\right\rangle$, respectively.

Refs. \cite{pair-o6,mcsm} already provided \begin{equation}\label{eq-ove-1} \left\langle\gamma_1,N\right\rangle=-\frac{1}{2}N\sum_{l=0}^{N-1}tr(\gamma_1\lambda^{2l+1})J^{N-1}_l,\\ \end{equation} where the $J$ tensor is defined in Eq. (\ref{eq-J-tensor}).

\begin{widetext}

Now we turn to the more complicated overlap $\left\langle \gamma_1,\gamma_2,N\right\rangle$. It is given by \begin{equation} \begin{aligned} &\left\langle \gamma_1,\gamma_2,N\right\rangle=\langle \Gamma_1\Gamma_2(\Lambda)^{N-2}(\Lambda^{\dagger})^N\rangle=\\ &\sum_{l=0}^{N-1}\left\langle \Gamma_1\left(\Lambda\right)^{N-2}\left(\Lambda^{\dagger}\right)^l\left[\Gamma_2,\Lambda^{\dagger}\right]\left(\Lambda^{\dagger}\right)^{N-1-l}\right\rangle\\ &=-\frac{1}{2}Ntr(\gamma_2\lambda)\left\langle \gamma_1,N-1\right\rangle\\ &+N(N-1)\left\langle \Gamma_1\left(\Lambda\right)^{N-2}\left(\Lambda^{\dagger}\right)^{N-2}\Gamma^{020\dagger}\right\rangle, \end{aligned} \end{equation} where the $\Gamma^{020}$ pair has a coefficient matrix $\gamma^{020}=\lambda\gamma_2\lambda$.

Next we further expand the last term of above equation according to \begin{equation} \begin{aligned} &\left\langle \Gamma_1\left(\Lambda\right)^{N-2}\left(\Lambda^{\dagger}\right)^{N-2}\Gamma^{020\dagger}\right\rangle=\left\langle \left(\Lambda\right)^{N-2}\left(\Lambda^{\dagger}\right)^{N-2}\left[\Gamma_1,\Gamma^{020\dagger}\right]\right\rangle\\ &+\sum_{l=0}^{N-3}\left\langle\left(\Lambda\right)^{N-2}\left(\Lambda^{\dagger}\right)^{l}\left[\Gamma_1,\Lambda^{\dagger}\right]\left(\Lambda^{\dagger}\right)^{N-l-3}\Gamma^{020\dagger}\right\rangle\\ &=-\frac{1}{2}tr(\gamma_1\lambda\gamma_2\lambda)I^{N-2}-\frac{1}{2}(N-2)tr(\gamma_1\lambda)\left\langle \lambda\gamma_{2}\lambda,N-2\right\rangle+(N-2)(N-3)\left\langle \lambda\gamma_{1}\lambda,\lambda\gamma_{2}\lambda,N-2\right\rangle\\ &+(N-2)\left\langle \lambda\gamma_1\lambda\gamma_2\lambda+\lambda\gamma_2\lambda\gamma_1\lambda,N-2\right\rangle. \end{aligned} \end{equation}

Combining the above two equations, we find that \begin{equation} \begin{aligned} &\left\langle \gamma_1,\gamma_2,N\right\rangle=-\frac{1}{2N(N-1)}tr(\gamma_1\lambda\gamma_2\lambda)J^{N}_2+\frac{1}{4}N(N-1)tr(\gamma_2\lambda)\sum_{l=0}^{N-2}tr(\gamma_1\lambda^{2l+1})J^{N-2}_l\\ &+\frac{1}{4}N(N-1)(N-2)^2tr(\gamma_1\lambda)\times\sum_{l=0}^{N-3}tr(\gamma_2\lambda^{2l+3})J^{N-3}_l-N(N-1)(N-2)^2\sum_{l=0}^{N-3}tr(\gamma_1\lambda\gamma_2\lambda^{2l+3})J^{N-3}_l\\ &+N(N-1)(N-2)(N-3)\left\langle\lambda\gamma_{1}\lambda,\lambda\gamma_{2}\lambda,N-2\right\rangle ~. \end{aligned} \end{equation}

By solving the above recursion relation, we can express the $\left\langle \gamma_1,\gamma_2,N\right\rangle$ overlap in terms of matrix traces and the overlap-related $J$ tensor as

\begin{equation} \begin{aligned} &\langle \gamma_1,\gamma_2,N\rangle=\langle \Gamma_1\Gamma_2(\Lambda)^{N-2}(\Lambda^{\dagger})^N\rangle\\ &=-\frac{1}{2}N(N-1)\sum_{k=0}^{\frac{N-2-N\%2}{2}}\left[\frac{(N-2)!!(N-3)!!}{(N-2k)!!(N-2k-1)!!}\right]^2tr(\gamma_1\lambda^{2k+1}\gamma_2\lambda^{2k+1})J^{N-2k}_2\\ &+\frac{1}{4}N(N-1)\sum_{k=0}^{\frac{N-2-N\%2}{2}}\left[\frac{(N-2)!!(N-3)!!}{(N-2k-2)!!(N-2k-3)!!}\right]^2tr(\gamma_2\lambda^{2k+1})\sum_{l=0}^{N-2k-2}tr(\gamma_1\lambda^{2l+2k+1})J^{N-2k-2}_l\\ &+\frac{1}{4}N(N-1)\sum_{k=0}^{\frac{N-3+N\%2}{2}}\left[\frac{(N-2)!!(N-3)!!}{(N-2k-3)!!(N-2k-4)!!}\right]^2tr(\gamma_1\lambda^{2k+1})\sum_{l=0}^{N-2k-3} tr(\gamma_2\lambda^{2k+2l+3})J^{N-2k-3}_l\\ &-N(N-1)\sum_{k=0}^{\frac{N-3+N\%2}{2}}\left[\frac{(N-2)!!(N-3)!!}{(N-2k-3)!!(N-2k-4)!!}\right]^2\sum^{N-2k-3}_{l=0} tr\left(\gamma_1\lambda^{2k+1}\gamma_2\lambda^{2k+2l+3}\right)J^{N-2k-3}_l. \end{aligned} \end{equation} where the $\%$ symbol is a remainder operator as in the C-language standard.

Following the same philosophy, we express $\left\langle \gamma_1,\gamma_2,\gamma_3,N\right\rangle$ as \begin{equation} \begin{aligned} &\langle \gamma_1,\gamma_2,\gamma_3,N\rangle=\langle \Gamma_1\Gamma_2\Gamma_3(\Lambda)^{N-3}(\Lambda^{\dagger})^N\rangle\\ &=\frac{1}{N(N-1)(N-2)}\sum_{t=0}^{\frac{N-3-N\%3}{3}}\left[\frac{N!!!(N-1)!!!(N-2)!!!}{(N-3t)!!!(N-3t-1)!!!(N-3t-2)!!!}\right]^2\\ &\times\left[\frac{1}{4}tr(\gamma_2\lambda^{2t+1})tr(\gamma_1\lambda^{2t+1}\gamma_3\lambda^{2t+1})-tr(\gamma_1\lambda^{2t+1}\gamma_2\lambda^{2t+1}\gamma_3\lambda^{2t+1})\right]J^{N-3t}_3\\ &+N(N-1)(N-2)\sum_{t=0}^{\frac{N-3-N\%3}{3}-(N\%3==0)}\frac{1}{N-3t-3}\left[\frac{(N-3)!!!(N-4)!!!(N-5)!!!}{(N-3t-6)!!!(N-3t-4)!!!(N-3t-5)!!!}\right]^2\\ &\times \left\langle \tilde \gamma,N-3-3t\right\rangle\\ &-\frac{1}{2}N(N-1)(N-2)\sum_{t=0}^{\frac{N-3-N\%3}{3}}\frac{1}{N-3t-2}\left[\frac{(N-3)!!!(N-4)!!!(N-5)!!!}{(N-3t-3)!!!(N-3t-4)!!!(N-3t-5)!!!}\right]^2\\ &\times tr(\gamma_3\lambda^{2t+1}\gamma_2\lambda^{2t+1})\left\langle \lambda^t\gamma_1\lambda^t,N-2-3t\right\rangle\\ &-\frac{1}{2}\sum_{t=0}^{\frac{N-3-N\%3}{3}}\frac{N!!!(N-1)!!!(N-2)!!!(N-3)!!!(N-4)!!!(N-5)!!!}{\left[(N-3t-3)!!!\right]^2(N-3t-1)!!!(N-3t-2)!!!(N-3t-4)!!!(N-3t-5)!!!}\\ &\times tr(\gamma_3\lambda^{2t+1})\left\langle \lambda^t\gamma_1\lambda^t,\lambda^t\gamma_2\lambda^t,N-3t-1\right\rangle\\ &+\sum_{t=0}^{\frac{N-3-N\%3}{3}-(N\%3==0||N\%3==1)}\frac{N!!!(N-1)!!!(N-2)!!!(N-3)!!!(N-4)!!!(N-5)!!!}{(N-3t-3)!!!(N-3t-4)!!!\left[(N-3t-5)!!!\right]^2(N-3t-6)!!!}\\ &\times\frac{1}{(N-3t-7)!!!}\widetilde{\sum\left\langle\cdots\right\rangle},\\ \end{aligned} \end{equation} where the boolean expressions $(N\%3==0)$ and $(N\%3==0||N\%3==1)$ are equal to 0 or 1, following the C-language standard, the $\tilde \gamma$ is the following sum of five skew-symmetric matrices, \begin{equation*} \begin{aligned} &\tilde \gamma=\left[\frac{1}{4}tr(\gamma_1\lambda^{2t+1})tr(\gamma_2\lambda^{2t+1})-\frac{1}{2}tr(\gamma_1\lambda^{2t+1}\gamma_2\lambda^{2t+1})\right]\lambda^{t+1}\gamma_3\lambda^{t+1}\\ &-\frac{1}{2}tr(\gamma_2\lambda^{2t+1})\lambda^{t+1}\left[\gamma_1\lambda^{2t+1}\gamma_3+\gamma_3\lambda^{2t+1}\gamma_1\right]\lambda^{t+1}+\lambda^{t+1}\left[\gamma_2\lambda^{2t+1}\gamma_1\lambda^{2t+1}\gamma_3+\gamma_3\lambda^{2t+1}\gamma_1\lambda^{2t+1}\gamma_2\right]\lambda^{t+1}\\ &-\frac{1}{2}tr(\gamma_1\lambda^{2t+1}\gamma_3\lambda^{2t+1})\lambda^{t+1}\gamma_2\lambda^{t+1}-\frac{1}{2}tr(\gamma_1\lambda^{2t+1})\lambda^{t+1}\left[\gamma_2\lambda^{2t+1}\gamma_3+\gamma_2\lambda^{2t+1}\gamma_3\right]\lambda^{t+1}\\ &+\lambda^{t+1}\left[\gamma_1\lambda^{2t+1}\gamma_2\lambda^{2t+1}\gamma_3+\gamma_1\lambda^{2t+1}\gamma_3\lambda^{2t+1}\gamma_2+\gamma_2\lambda^{2t+1}\gamma_3\lambda^{2t+1}\gamma_1+\gamma_3\lambda^{2t+1}\gamma_2\lambda^{2t+1}\gamma_1\right]\lambda^{t+1},\\ \end{aligned} \end{equation*} and $\widetilde{\sum\left\langle\cdots\right\rangle}$ is the following sum of five overlaps, \begin{equation*} \begin{aligned} &\widetilde{\sum\left\langle\cdots\right\rangle}=\sum_{i\neq j~j\neq k~i\neq k}\left\langle\lambda^{t+1}\gamma_i\lambda^{t+1},\lambda^{t+1}\{\gamma_j\lambda^{2t+1}\gamma_k+\gamma_k\lambda^{2t+1}\gamma_j\}\lambda^{t+1},N-3t-3\right\rangle\\ &-\frac{1}{2}tr(\gamma_2\lambda^{2t+1})\left\langle \lambda^{t+1}\gamma_1\lambda^{t+1},\lambda^{t+1}\gamma_3\lambda^{t+1},N-3t-3\right\rangle-\frac{1}{2}tr(\gamma_1\lambda^{2t+1})\left\langle\lambda^{t+1}\gamma_2\lambda^{t+1},\lambda^{t+1}\gamma_3\lambda^{t+1},N-3t-3\right\rangle.\\ \end{aligned} \end{equation*}

For the matrix element of an arbitrary one-body operator $Q$, we have \begin{equation}\label{eq-q} \begin{aligned} &\left\langle\left(\Lambda\right)^{N}Q\left(\Lambda^{\dagger}\right)^N\right\rangle=\sum_{l=0}^{N-1}\left\langle\left(\Lambda\right)^{l}\left[\Lambda,Q\right]\left(\Lambda\right)^{N-l-1}(\Lambda^{\dagger})^N\right\rangle=N\left\langle \lambda q+q^T\lambda,N\right\rangle=-N^2\sum_{l=0}^{N-1}tr(q\lambda^{2l+2})J^{N-1}_{l}\\ \end{aligned} \end{equation} One sees that for any one-body operator, $\left\langle\left(\Lambda\right)^{N}Q\left(\Lambda^{\dagger}\right)^N\right\rangle=\left\langle\left(\Lambda\right)^{N}Q^{\dagger}\left(\Lambda^{\dagger}\right)^N\right\rangle$, where the coefficient matrix of $Q^{\dagger}$ is $q^T$, and $tr(q^T\lambda^{2l+2})=tr(q\lambda^{2l+2})$. This is as required for a general Hermitian operator.

For the matrix element of an arbitrary two-body operator $\Gamma^{\dagger}\Gamma$, where the $\Gamma$ pair has coefficient matrix $\gamma$, \begin{equation}\label{eq-pp} \begin{aligned} &\left\langle \left(\Lambda\right)^{N}\Gamma^{\dagger}\Gamma(\Lambda^{\dagger})^N\right\rangle=\sum_{l=0}^{N-1}\left\langle\left(\Lambda\right)^{l}\left[\Lambda,\Gamma^{\dagger}\right]\left(\Lambda\right)^{N-l-1}\Gamma \left(\Lambda^{\dagger}\right)^N\right\rangle=-\frac{1}{2}Ntr(\gamma\lambda)\left\langle\gamma,N\right\rangle+N(N-1)\left\langle\gamma,\lambda\gamma\lambda,N\right\rangle.\\ \end{aligned} \end{equation}

For an arbitrary operator $\hat O$, the derivative of its matrix element along the direction of an arbitrary $\Gamma$ pair reads \begin{equation}\label{eq-diff-detail} \begin{aligned} &\frac{\partial \left\langle\left(\Lambda\right)^{N}\hat O(\Lambda^{\dagger})^N\right\rangle}{\partial \delta_{\parallel\Gamma}}=\lim_{\delta_{\parallel\Gamma}\rightarrow 0}\frac{\left\langle\left(\Lambda+\delta_{\parallel\Gamma}\Gamma\right)^{N}\hat O\left(\Lambda^{\dagger}+\delta_{\parallel\Gamma}\Gamma^{\dagger}\right)^N\right\rangle-\left\langle \left(\Lambda\right)^{N}\hat O\left(\Lambda^{\dagger}\right)^N\right\rangle}{\delta_{\parallel\Gamma}}\\ &=\lim_{\delta_{\parallel\Gamma}\rightarrow 0}\frac{\delta_{\parallel\Gamma} N\left\langle\Gamma\left(\Lambda\right)^{N-1}\hat O(\Lambda^{\dagger})^N\right\rangle+\delta_{\parallel\Gamma} N\left\langle\left(\Lambda\right)^{N}\hat O(\Lambda^{\dagger})^{N-1}\Gamma^{\dagger}\right\rangle+O(\delta_{\parallel\Gamma}^2)}{\delta_{\parallel\Gamma}}\\ &=N\left\langle\Gamma\left(\Lambda\right)^{N-1}(\hat O+\hat O^{\dagger})(\Lambda^{\dagger})^N\right\rangle,\\ \end{aligned} \end{equation} where $O(\delta_{\parallel\Gamma}^2)$ is the second infinitely small quantity of $\delta_{\parallel\Gamma}$.

In our variation, we choose all the $\lambda_{ij}$ as our variables. With respect to a single matrix element $\lambda_{kl}$, the corresponding direction pair is $\Lambda^{kl}$ with coefficient matrix $\lambda^{kl}$ and associated matrix element $\lambda^{kl}_{ij}=\delta_{ik}\delta_{jl}-\delta_{jk}\delta_{il}$, where the $\delta$ is the usual Kronecker symbol.

If $\hat O$ is the identity operator, then Eq. (\ref{eq-diff}) reduces to the first derivative of the overlap $I^N$, namely
\begin{equation}\label{eq-diff-ove} \frac{\partial \left\langle\left(\Lambda\right)^{N}(\Lambda^{\dagger})^N\right\rangle}{\partial \lambda_{kl}}=2N\left\langle\lambda^{kl},N\right\rangle=2N^2\sum_{m=0}^{N-1}(\lambda^{2m+1})_{kl}J^{N-1}_m. \end{equation}

If $\hat O=Q$ is a one-body operator, then \begin{equation}\label{eq-diff-q} \begin{aligned} &\frac{\partial \left\langle\left(\Lambda\right)^{N}Q(\Lambda^{\dagger})^N\right\rangle}{\partial \lambda_{kl}}=N\left\langle\Lambda^{kl}(\Lambda)^{N-1}(Q+Q^{\dagger})(\Lambda^{\dagger})^N\right\rangle=N\left\langle\left[\Lambda^{kl},Q+Q^{\dagger}\right]\left(\Lambda\right)^{N-1}(\Lambda^{\dagger})^N\right\rangle\\ &+N\sum_{m=0}^{N-2}\left\langle\Lambda^{kl}\left(\Lambda\right)^m\left[\Lambda,Q+Q^{\dagger}\right]\left(\Lambda\right)^{N-m-2}(\Lambda^{\dagger})^N\right\rangle=N\left\langle \tilde \gamma^{kl},N\right\rangle+N(N-1)\left\langle \lambda^{kl},\tilde \gamma,N\right\rangle, \end{aligned} \end{equation} with $\tilde\gamma^{kl}=\lambda^{kl}(q+q^T)+(q+q^T)\lambda^{kl},~\tilde\gamma=\lambda(q+q^T)+(q+q^T)\lambda$.

If $\hat O=\Gamma^{\dagger}\Gamma$ is a two-body operator, then \begin{equation}\label{eq-diff-pp} \begin{aligned} &\frac{\partial \left\langle\left(\Lambda\right)^{N}\Gamma^{\dagger}\Gamma(\Lambda^{\dagger})^N\right\rangle}{\partial \lambda_{kl}}=2N\left\langle\Lambda^{kl}(\Lambda)^{N-1}\Gamma^{\dagger}\Gamma(\Lambda^{\dagger})^N\right\rangle=2N\left\langle\left[\Lambda^{kl},\Gamma^{\dagger}\right]\left(\Lambda\right)^{N-1}\Gamma(\Lambda^{\dagger})^N\right\rangle\\ &+2N\sum_{m=0}^{N-2}\left\langle\Lambda^{kl}\left(\Lambda\right)^m\left[\Lambda,\Gamma^{\dagger}\right]\left(\Lambda\right)^{N-m-2}\Gamma(\Lambda^{\dagger})^N\right\rangle=-Ntr(\lambda^{kl}\gamma)\left\langle\gamma,N\right\rangle-N(N-1)tr(\lambda\gamma)\left\langle\lambda^{kl},\gamma,N\right\rangle\\ &+2N(N-1)(N-2)\left\langle\lambda^{kl},\lambda\gamma\lambda,\gamma,N\right\rangle+2N(N-1)\left\langle\lambda^{kl}\gamma\lambda+\lambda\gamma\lambda^{kl},\gamma,N\right\rangle\\ &=2N\gamma_{kl}\left\langle\gamma,N\right\rangle+2N(N-1)\left\langle \lambda^{kl}\gamma\lambda+\lambda\gamma\lambda^{kl}-\frac{1}{2}tr(\lambda\gamma)\lambda^{kl},\gamma,N\right\rangle+2N(N-1)(N-2)\left\langle\lambda^{kl},\lambda\gamma\lambda,\gamma,N\right\rangle \end{aligned} \end{equation}

\end{widetext}


\begin{thebibliography}{10}
\bibitem{sm-1} M. G. Mayer, Phys. Rev. {\bf 75}, 1969 (1949).

\bibitem{sm-2} J. H. D. Jensen, $et~al.$, Naturwissenschaften {\bf 36}, 155 (1949).

\bibitem{npa-phys-rep} Y. M. Zhao and A. Arima, Phys. Rep. {\bf 545}, 1 (2014).

\bibitem{rev-ibm} F. Iachello and I. Talmi, Rev. Mod. Phys. {\bf 59}, 339 (1987).

\bibitem{npa-cal-sd1} Y. A. Luo, $et~al.$, Nucl. Phys. {\bf A 669}, 101 (2000).

\bibitem{npa-cal-sd2} Y. M. Zhao, S. Yamaji, N. Yoshinaga, and A. Arima, Phys. Rev. {\bf C 62},
014315 (2000).

\bibitem{npa-cal-sd4} L. Y. Jia, H. Zhang, and Y. M. Zhao, Phys. Rev. {\bf C 75}, 034307 (2007).

\bibitem{npa-cal-sd6} X. F. Meng, F. R. Wang, Y. A. Luo, F. Pan, and J. P. Draayer, Phys. Rev. {\bf C
77}, 047304 (2008).

\bibitem{npa-cal-other1} G. A. Jones, P. H. Regan, Z. Podolyak, N. Yoshinaga, K. Higashiyama, G.
deAngelis, Y. H. Zhang, A. Gadea, C. A. Ur, $et~al.$, Phys. Rev. {\bf C 76}, 054317 (2007).

\bibitem{npa-cal-other3} Z. Y. Xu, Y. Lei, Y. M. Zhao, S. W. Xu, Y. X. Xie, and A. Arima, Phys, Rev.
{\bf C 79}, 054315 (2009).

\bibitem{npa-validity-1} Y. Lei, Z. Y. Xu, Y. M. Zhao, and A. Arima, Phys. Rev. {\bf C 80}, 064316
(2009).
\bibitem{npa-validity-2} Y. Lei, Z. Y. Xu, Y. M. Zhao, A. Arima, Phys. Rev. {\bf C 82}, 034303
(2010).
\bibitem{npa-validity-3} Y. Lei, Y. M. Zhao, A. Arima, Phys. Rev. {\bf C 84}, 044301 (2011).

\bibitem{lifetime} A. Dewald, S. Harissopulos, G. Bohm, A. Gelberg, K. P. Schmittgen, R. Wirowski, K.
O. Zell, and P. vonBrentano, Phys. Rev. {\bf C 37}, 289 (1988).

\bibitem{gfactor1} S. Harissopulos, A. Gelberg, A. Dewald, M. Hass, L. Weissman, and C. Broude, Phys.
Rev. {\bf C 52}, 1796
(1995).

\bibitem{gfactor2} P. Das, R. G. Pillay, V. V. Krishnamurthy, S. N. Mishra, and S. H. Devare, Phys.
Rev. {\bf C 53}, 1009 (1996).

\bibitem{yoshinaga-ba} K. Higashiyama, N. Yoshinaga, K. Tanabe, Phys. Rev. {\bf C 67}, 044305
(2003).

\bibitem{lei-ba} Y. Lei and Z. Y. Xu, Phys. Rev. {\bf C 92}, 014317 (2015).

\bibitem{cheng-ba} Y. Y. Cheng, Y. Lei, Y. M. Zhao, and A. Arima, Phys. Rev. {\bf C 92}, 064320
(2015).

\bibitem{hfb-1} K. Sugawara-Tanabe and A. Arima, Phys. Lett. {\bf B 110}, 87 (1982).

\bibitem{hfb-2} S. Pittel and J. Dukelsky, Phys. Lett. {\bf B 128}, 9 (1983).

\bibitem{lei-n74} Y. Lei, G. J. Fu, and Y. M. Zhao, Phys. Rev. {\bf C 87}, 044331 (2013).

\bibitem{pair-o6} T. Mizusaki and T. Otsuka, Prog. Theo. Phys. Supp. {\bf 125}, 97 (1996).

\bibitem{mcsm} T. Otsuka, M. Honma, T. Mizusaki, N. Shimizu, and Y. Utsuno, Prog. Part. Nucl. Phys.
{\bf 47}, 319 (2001).

\bibitem{blas} http://www.netlib.org/blas.

\bibitem{bfgs-1} C. G. Broyden, J. Inst. Math. Appl. {\bf 6}, 76 (1970).

\bibitem{bfgs-2} R. Fletcher, Comp. J. {\bf 13}, 317 (1970).

\bibitem{bfgs-3} D. Goldfarb, Math. Comp. {\bf 24}, 23 (1970).

\bibitem{bfgs-4} D. F. Shanno, Math. Comp. {\bf 24}, 647 (1970).

\bibitem{npa-for-1} J. Q. Chen, Nucl. Phys. {\bf A 626}, 686 (1997).

\bibitem{npa-for-2} Y. M. Zhao, N. Yoshinaga, S. Yamaji, J. Q. Chen, and A. Arima, Phys. Rev. {\bf C
62}, 014304 (2000).

\bibitem{smmc} J. A. White, S. E. Koonin, and D. J. Dean, Phys. Rev {\bf C 61}, 034303 (2000).

\bibitem{pbcs} Y. Lei, Z. Y. Xu, Y. M. Zhao, and A. Arima, Phys. Rev. {\bf C 80}, 064316 (2009).

\bibitem{ensdf} http://www.nndc.bnl.gov/ensdf/.

\bibitem{in-1} D. R. Inglis, Phys. Rev. 96, 1059 (1954).

\bibitem{in-2} D. R. Inglis, Phys. Rev. 103, 1786 (1956).

\bibitem{crank-review-1} M. J. A. de Voigt, J. Dudek, and Z. Szyma\'nski, Rev. Mod. Phys., 55, 949
(1983).

\bibitem{crank-review-2} A. V. Afanasjev, D. B. Fossan, G. J. Lane, and I. Ragnarsson, Phys. Rep.
322,1 (1999).

\bibitem{crank-review-3} S. Frauendorf, Rev. Mod. Phys. 73, 463 (2001).

\bibitem{crank-ba-1} E. S. Paul, D. B. Fossan, Y. Liang, R. Ma, and N. Xu, Phys. Rev. C, 40, 1255 (1989).

\bibitem{crank-ba-2} S. Juutinen, S. T\"orm\"anen, P. Ahonen, M. Carpenter, C. Fahlander, J. Gascon, $et~al.$, Phys. Rev. C 52, 2946
(1995).

\bibitem{val-2} Y. Lei, Z. Y. Xu, Y. M. Zhao, and A. Arima, Phys. Rev. {\bf C 82}, 034303 (2010).

\bibitem{val-3} Y. Lei, Y. M. Zhao, and A. Arima, Phys. Rev. {\bf C 84}, 044301 (2011).

\bibitem{sherpa} I. Stetcu and C. W. Johnson, Phys. Rev. {\bf C 66}, 034301(2002); {\it ibid.} {\bf
67}, 044315 (2003); {\it ibid.} {\bf 69}, 024311 (2004).
\end{thebibliography}
\end{document}